\documentclass[11pt,a4paper]{article}

\usepackage{hyperref}  

\usepackage{cite}

\usepackage[T1]{fontenc}    
\usepackage{lmodern}
\usepackage{times}

\usepackage{pgf}
\usepackage{tikz}

\usepackage{doi}

\usepackage[fleqn]{amsmath}
\usepackage{amsfonts,amsthm,amssymb}

\numberwithin{equation}{section}

\newtheorem{theorem}{Theorem}[section]
\newtheorem{identity}{Identity}
\newtheorem{prop}{Proposition}[section]

{\theoremstyle{remark}
\newtheorem{rem}{\sl Remark}

\newcommand{\bra}[1]{\langle\,#1\,|}
\newcommand{\ket}[1]{|\,#1\,\rangle}

\newcommand{\moy}[1]{\langle\,#1\,\rangle}

\def\tr{\operatorname{tr}}

\DeclareMathSymbol{\Alpha}{\mathalpha}{operators}{"41}
\DeclareMathSymbol{\Beta}{\mathalpha}{operators}{"42}
\DeclareMathSymbol{\Epsilon}{\mathalpha}{operators}{"45}
\DeclareMathSymbol{\Zeta}{\mathalpha}{operators}{"5A}
\DeclareMathSymbol{\Eta}{\mathalpha}{operators}{"48}
\DeclareMathSymbol{\Iota}{\mathalpha}{operators}{"49}
\DeclareMathSymbol{\Kappa}{\mathalpha}{operators}{"4B}
\DeclareMathSymbol{\Mu}{\mathalpha}{operators}{"4D}
\DeclareMathSymbol{\Nu}{\mathalpha}{operators}{"4E}
\DeclareMathSymbol{\Omicron}{\mathalpha}{operators}{"4F}
\DeclareMathSymbol{\Rho}{\mathalpha}{operators}{"50}
\DeclareMathSymbol{\Tau}{\mathalpha}{operators}{"54}
\DeclareMathSymbol{\Chi}{\mathalpha}{operators}{"58}
\DeclareMathSymbol{\omicron}{\mathord}{letters}{"6F}

\title{On scalar products and form factors by Separation of Variables: the antiperiodic XXZ model}
\usepackage{authblk}\author[1]{Hao Pei\footnote{hao.pei@universite-paris-saclay.fr}}\author[1]{V\'{e}ronique Terras\footnote{veronique.terras@universite-paris-saclay.fr}}\affil[1]{Université Paris-Saclay, CNRS,  LPTMS, 91405, Orsay, France}
\setcounter{Maxaffil}{0}

\begin{document}

\maketitle

\begin{abstract} 
We consider the XXZ spin-1/2 Heisenberg chain with antiperiodic boundary conditions. The inhomogeneous version of this model can be solved by Separation of Variables (SoV), and the eigenstates can be constructed in terms of $Q$-functions, solution of a Baxter TQ-equation, which have double periodicity compared to the periodic case. We compute in this framework the scalar products of a particular class of separate states which notably includes the eigenstates of the transfer matrix. We also compute the form factors of local spin operators, i.e. their matrix elements between two eigenstates of the transfer matrix. We show that these quantities admit determinant representations with rows and columns labelled by the roots of the $Q$-functions of the corresponding separate states, as in the periodic case, although the form of the determinant are here slightly different. We also propose alternative types of determinant representations written directly in terms of the transfer matrix eigenvalues.
\end{abstract}

\tableofcontents

\section{Introduction}

The exact determination of the spectrum of physical quantum models is a very old problem which comes back to the seminal work of Bethe \cite{Bet31} concerning the diagonalization of the Heisenberg spin chain Hamiltonian by the method now known as {\em Bethe Ansatz}. In the late seventies, an algebraic framework to tackle this problem was settled with the development of the {\em Quantum Inverse Scattering Method} (QISM) \cite{FadST79,FadT79}, and the models for which the method applies were called {\em quantum integrable models}. There exist now a wide range of models for which one knows how to exactly characterize the spectrum and eigenstates of the Hamiltonian, with applications in various area of physics, from condensed matter to high energy physics.

The computation of correlation functions of such quantum integrable models is {\em a priori} a much more difficult problem: the first exact results concerning this problem \cite{JimMMN92,JimM95L,JimM96,KitMT99,KitMT00,KitMST02a} were obtained more than sixty years after the work of Bethe. In particular, an important progress was made in \cite{KitMT99}, where determinant representations for the form factors, i.e. the matrix elements of the local operators in the eigenstates of the Hamiltonian, were obtained for the periodic XXZ spin chain in finite volume within the {\em Algebraic Bethe Ansatz} (ABA) framework \cite{Fad96}. Such determinant representations could be derived from the fact that, on the one hand, there exist a similar determinant representation  for the scalar products of Bethe states \cite{Sla89}, and that, on the other hand, one can explicitly reconstruct the local spin operators as some matrix element of the monodromy matrix dressed by a product of transfer matrices \cite{KitMT99,MaiT00,GohK00}, hence enabling the computation of their action on the Bethe states in a simple way. Such determinant representations for the form factors proved to be very useful for the computation of correlation functions and structure factors, either numerically \cite{BieKM02,BieKM03,CauM05,CauHM05} or analytically \cite{KitMST05a,KitMST05b,KitKMST09b,KitKMST09c,KitKMST11a,KozT11,Koz11,KitKMST11b,KitKMST12,KitKMT14,DugGK13,DugGK14,GohKKKS17,Koz18a}.
See also \cite{BooJMST05,BooJMST06,BooJMST06a,BooJMST06b,BooJMST06c,BooJMST07,BooJMST09,JimMS09,JimMS11} for an alternative approach to the problem of the computation of correlation functions.

It is worth mentioning that the aforementioned results concern essentially the simplest integrable models solvable by ABA, namely the XXZ Heisenberg spin chain or the quantum non-linear Schrödinger model with periodic boundary conditions (see also  \cite{KitKMNST07,KitKMNST08,LevT13a,LevT14a,GriDT19} for some results about correlation functions in more complicated models within ABA, and \cite{BelPRS12b,HutLPRS18,PakRS18,BelS19,SlaZZ20} for some generalizations of the scalar products and/or form factor determinant representations to more complicated models). For quantum integrable models not solvable within ABA but within the quantum version of the {\em Separation of Variables} (SoV) approach \cite{Skl85,Skl90,Skl92,Skl95}, the problem of computing correlation functions still remains widely open. It has yet be shown that this method was very powerful, in that it could be applied to characterize the spectrum and eigenstates of a large class of quantum integrable models \cite{BabBS96,Smi98a,DerKM01,DerKM03,DerKM03b,BytT06,vonGIPS06,FraSW08,AmiFOW10,NicT10,Nic10a,Nic11,FraGSW11,GroMN12,GroN12,Nic12,Nic13,Nic13a,Nic13b,GroMN14,FalN14,FalKN14,NicT15,LevNT16,NicT16,JiaKKS16,GroLS17,MaiNP17,MaiN18,RyaV19,MaiN19,MaiN19b,MaiN19c}.


One of the problems, for the computation of physical quantities such as correlation functions within the SoV approach, is that  this method requires in general to deform the model by introducing non-physical inhomogeneity parameters. The latter ensure the definition of a basis of the space of states, the SoV basis, in which the spectral problem for the transfer matrix can be conveniently addressed.  However, the solution of this spectral problem is usually given in terms of a system of discrete equations that need to be transformed to more conventional ones (typically in the form of functional TQ-equations involving the transfer matrix eigenvalues and a Baxter $Q$-function of a particular analytical form \cite{Bax82L}) for the consideration of the physical model at the homogeneous limit, i.e. when all inhomogeneity parameters coincide. This may be completely straightforward, as in the XXX spin chain with antiperiodic (or twisted) boundary conditions \cite{KitMNT16}: in that case, the $Q$-function is simply a polynomial, leading to a characterization of the spectrum in terms of Bethe equations similar in their form to those obtained by ABA in the periodic case, and the transformation from the discrete to the continuous characterization of the spectrum follows from a mere polynomial interpolation. Even for simple models such as Heisenberg spin $1/2$ chains, this may however be more complicated. For instance, in the antiperiodic XXZ spin chain that we consider in the present article, the $Q$-function is no longer a usual trigonometric polynomial (or a Laurent polynomial in the exponential variable) as in the periodic case, but a trigonometric polynomial with double period \cite{BatBOY95,NicT15}. And for open XXZ spin chains with completely generic boundary fields, the problem of finding a good description of the spectrum in terms of a functional TQ-equation is still open \cite{KitMN14}, at least if one restricts oneself to usual TQ-equations (it is indeed possible in that case to describe the SoV spectrum  in terms of a TQ equation with an additional inhomogeneous term, see also \cite{WanYCS15L} or \cite{BelC13,Bel15,BelP15,AvaBGP15}).

Once the spectral problem has been solved, the next goal towards correlation functions is, as in the ABA approach, the computation of the form factors of local operators. This means on the one hand that we should act with local operators on the transfer matrix eigenstates: as in the ABA case, this is possible via an adaptation of the solution of the quantum inverse problem \cite{KitMT99,MaiT00,GohK00} to the model that one considers. In general, the result of such an action should be a sum over a particular sub-class of {\em separate states}, a class of states which includes (but is more general than) the transfer matrix eigenstates. On the other hand, one should be able to compute the scalar products of the resulting separate states in a convenient way, as in the ABA approach \cite{Sla89}. 

It happens that, for a large class of quantum integrable models solved by SoV, the scalar products of the separate states can be generically expressed as determinants of a dressed sum of generalized Vandermonde matrices. Such determinants depend however in a non-trivial way on the inhomogeneity parameters (the latter label the rows and columns), and cannot be easily used for the study of the physical model. 
In simple models such as the XXX spin chain, for which the Baxter $Q$-function is a usual polynomial, it has been recently shown \cite{KitMNT16} (see also \cite{KitMNT17,KitMNT18}) that it was possible to transform these determinants into more convenient ones, of Izergin \cite{Ize87} or Slavnov \cite{Sla89} types: in the latter case, the rows and columns of the determinants are notably directly labelled by the roots of the involved $Q$-functions (i.e. the Bethe roots in the case of an eigenstate). In particular, in the antiperiodic (or twisted) XXX spin chain, it is possible to compute the form factors \cite{KitMNT16} and the correlation functions \cite{NicPT20} directly in the SoV approach, and to recover the same results as from ABA \cite{KitMT99,KitMT00}. However, it should be noticed that the XXX spin chain is in this context a particularly simple example and that the method proposed in \cite{KitMNT16,KitMNT17,KitMNT18} is not generic enough to apply to more complicated models. In fact, this method breaks down when the Baxter $Q$-function is no longer a usual polynomial (or trigonometric/Laurent polynomial) in the variables of the model, or in other terms, when the eigenstates cannot be simply represented as usual Bethe states\footnote{i.e. as the multiple action of an operator entry of the monodromy matrix (or of an operator which is a linear combination of such entries) on a given reference state.}. Nor seems the alternative method proposed in \cite{BelS19} directly applicable in that case.

It is therefore very important, for the applicability of the SoV approach to the computation of correlation functions in more complicated and physically interesting models, to understand how to deal with such cases.  In this respect, the consideration of the antiperiodic (or twisted antiperiodic) XXZ chain provides an interesting framework for the development of a more generic approach to the scalar products and form factors within the SoV approach: indeed, in that case the  Baxter $Q$-function is not a mere polynomial and, as already mentioned, the method used in \cite{KitMNT16,KitMNT17,KitMNT18} breaks down.

We explain here how it is possible to transform the determinants issued from SoV in this case.
More precisely, we consider the scalar product of a sub-class of separate states which notably contains the transfer matrix eigenstates. As in \cite{KitMNT16,KitMNT17,KitMNT18}, we show that the latter can be transformed into generalized versions of Izergin determinants. However, due to the double periodicity of the $Q$-functions indexing the considered separate states, and contrary to what happened in \cite{KitMNT16,KitMNT17,KitMNT18}, these generalized Izergin determinants still depend in a non-trivial way on the inhomogeneity parameters.
So as to eliminate this dependance, we multiply the involved matrices by an adequately chosen matrix $\mathcal{X}$. We show that the resulting determinant present some similarities in its form with the so-called Slavnov determinant obtained in the periodic case \cite{Sla89}, with however some noticeable differences related to the analytical differences of the two cases. We also show that our transformation can be adapted to compute the form factors of local operators, which can be then represented in terms of a determinant of a sum of two matrices: the matrix of the scalar product and a rank 1 matrix. We finally show that it is also possible to represent the overlaps of two twisted eigenstates, as well as the form factors, directly in terms of the transfer matrix eigenvalues. 

We would like to underline that we deal here with a description of the transfer matrix spectrum in terms of solutions a usual TQ-equation, i.e. homogeneous in the terminology of \cite{NicT15}. Indeed, as shown in  \cite{NicT15} for the antiperiodic XXZ model, it is also possible to describe the transfer matrix spectrum in the context of SoV in terms of a TQ-equation with an extra term, called {\em inhomogeneous} TQ-equation, so as to deal with simpler types of $Q$-functions. This approach was notably proposed in the context of the so-called {\em off-diagonal Bethe Ansatz} (see \cite{WanYCS15L} and references therein). Such kind of description of the spectrum may also appear in the context of {\em modified Bethe Ansatz} \cite{BelC13,Bel15,BelP15,AvaBGP15,BelP16,BelSV18,BelPS21}. The advantage of this approach is the much simpler form of the $Q$-function (a usual trigonometric polynomial in the antiperiodic XXZ chain, see \cite{NicT15}), hence leading a priori, in the SoV context,  to  simpler algebraic transformations for the scalar products, as in \cite{KitMNT16}. The disadvantage is that the presence of the inhomogeneous term in the resulting Bethe equations complicates considerably the analysis of the configurations of the Bethe roots (see for instance \cite{BelF18}), and that it may therefore be a difficult task to analyse and control the obtained formulas for the ground states, low energy states, or other types of eigenstates, in the thermodynamic limit. 
Instead, we expect the studies of \cite{KitKMST09c,KitKMST11a} to be more easily transposable to the formulas that we obtain in the present paper.

The article is organized as follows. In Section~\ref{sec-XXZ} we introduce the model in the QISM framework. In Section~\ref{sec-SOV} we recall its solution by the quantum version of the Separation of Variables approach, i.e. the characterization of the transfer matrix spectrum and eigenstates. In Section \ref{sec-results}, we formulate our main results: the determinant representations for the scalar products of separate states and for the form factors of local spin operators.
The explicit computations leading to these results are then presented in Sections~\ref{sec-proof-SP} and \ref{sec-proof-ff} respectively.

\section{The XXZ model in the QISM framework}
\label{sec-XXZ}

In this paper, we consider the XXZ Heisenberg spin 1/2 chain of length $N$, 
\begin{equation}\label{Ham}
  H=\sum_{n=1}^N\big[\sigma_n^x\sigma_{n+1}^x+\sigma_n^y\sigma_{n+1}^y+\Delta(\sigma_n^z\sigma_{n+1}^z-1)\big].
\end{equation}
with twisted (quasi-periodic) boundary conditions:
\begin{equation}\label{cond-limit}
   \sigma_{N+1}^\alpha=K\,\sigma_1^\alpha\, K^{-1}.
\end{equation}
Here, $\Delta$ is the anisotropy parameter, $K$ is a $2\times 2$ numerical invertible twist matrix which will be specified later, and $\sigma_n^\alpha$, $\alpha=x,y,z$, denote Pauli matrices acting at site $n$ of the chain, i.e. on the local quantum spin space $V_n\simeq\mathbb{C}^2$, the $2^N$-dimensional quantum space of states of the model being $\mathcal{H}=\otimes_{n=1}^N V_n$.

The R-matrix of the model corresponds to the trigonometric solution of the Yang-Baxter equation,
\begin{equation}\label{mat-R}
R(\lambda )=
\begin{pmatrix}
\sinh \left( \lambda +\eta \right) & 0 & 0 & 0 \\ 
0 & \sinh \lambda & \sinh \eta & 0 \\ 
0 & \sinh \eta & \sinh \lambda & 0 \\ 
0 & 0 & 0 & \sinh (\lambda +\eta )
\end{pmatrix} ,
\end{equation}
in which $\eta$ is related to the anisotropy parameter $\Delta$ as
\begin{equation}
  \Delta=\cosh\eta.
\end{equation}
The monodromy matrix of the inhomogeneous spin-$1/2$ XXZ model of length $N$ with inhomogeneity parameters $\xi_1,\ldots,\xi_N$ 
is defined as the following ordered product of R-matrices:
\begin{align}\label{mon}
  T_0(\lambda) 
  &= R_{0, N}(\lambda-\xi_N)\ldots R_{0,1}(\lambda-\xi_1)
  =\begin{pmatrix} A(\lambda) & B(\lambda) \\ C(\lambda) & D(\lambda) \end{pmatrix}_{[0]},
\end{align}
where the indices $1,\ldots,N$ denote action on the quantum spaces $V_1\simeq\mathbb{C}^2,\ldots,V_N\simeq\mathbb{C}^2$, whereas the index $0$ denotes action on the auxiliary space $V_0\simeq\mathbb{C}^2$.
Hence $A(\lambda),B(\lambda),C(\lambda),D(\lambda)$ act on the quantum space of states $\mathcal{H}$ of the model, and $T_0(\lambda)$ acts on $V_0\otimes\mathcal{H}$.
The monodromy matrix \eqref{mon} satisfies with \eqref{mat-R} the following quadratic relation on $V_0\otimes V_{0'}\otimes \mathcal{H}$:
\begin{equation}\label{RTT}
   R_{0 0'}(\lambda-\mu)\, T_0(\lambda)\, T_{0'}(\mu)=T_{0'}(\mu)\, T_0(\lambda)\, R_{00'}(\lambda-\mu).
\end{equation}
The operators $A(\lambda)$, $B(\lambda)$, $C(\lambda)$, $D(\lambda)$ satisfy commutation relations issued from \eqref{RTT}, hence providing a representation of the corresponding Yang-Baxter algebra. The latter admits a central element, the quantum determinant, which is given by:
\begin{align}
   {\det}_q T(\lambda)=a(\lambda)\, d(\lambda-\eta)
   &= A(\lambda)\, D(\lambda-\eta)- B(\lambda)\, C(\lambda-\eta)\nonumber\\
   &= D(\lambda)\, A(\lambda-\eta)- C(\lambda)\, B(\lambda-\eta), 
   \label{q-det}
\end{align}
with
\begin{equation}\label{a-d}
   a(\lambda)=\prod_{n=1}^N \sinh(\lambda-\xi_n+\eta),
   \qquad
   d(\lambda)=\prod_{n=1}^N \sinh(\lambda-\xi_n).
\end{equation}

The transfer matrix of the model with $K$-twisted  boundary conditions is defined as
\begin{equation}\label{twist-transfer}
   \mathcal{T}_K(\lambda)=\tr_0 [ K_0\, T_0(\lambda) ].
\end{equation}
Provided that $[R(\lambda), K\otimes K]=0$, it generates a one-parameter family of commuting operators:
\begin{equation}
   [ \mathcal{T}_K(\lambda), \mathcal{T}_K(\mu)]=0, \qquad \forall \lambda,\mu\in\mathbb{C}.
\end{equation}
In the homogeneous limit $\xi_1,\ldots,\xi_N\to\eta/2$, we recover the Hamiltonian \eqref{Ham} of the XXZ spin chain with twist $K$ as
\begin{equation}
   H=2\sin\eta\ \lim_{\xi_1,\ldots,\xi_N\to\eta/2}
   \mathcal{T}_K(\lambda)^{-1}\,\frac{d}{d\lambda}\mathcal{T}_K(\lambda)\Big|_{\lambda=\eta/2}+\text{constant}.
\end{equation}

When $K$ is the identity matrix (periodic boundary conditions), or more generally a diagonal matrix, the model can be solved by Bethe Ansatz \cite{Bet31,FadT79}, and the eigenstates of the transfer matrix are constructed as Bethe states, i.e. by multiple action of the $B$ operators on the reference state $\ket{0}$ with all the spins up. A convenient determinant representation for the scalar product of such Bethe states was obtained in \cite{Sla89}, and was used in \cite{KitMT99} to express the form factors of local operators also as determinants. We will not consider this case in the present paper.

We are instead interested here in the case where $K$ is the $\sigma^x$ Pauli matrix (antiperiodic boundary conditions), or more generally the product of a diagonal matrix by $\sigma^x$ ($\kappa$-twisted antiperiodic boundary conditions),
\begin{equation}\label{form-K}
   K=\mathrm{diag}(\kappa, \kappa^{-1})\cdot \sigma^x, 
   \qquad \kappa \in \mathbb{C}\setminus\{0\}.
\end{equation}
In that case, usual Bethe Ansatz can no longer be used to construct the eigenstates of the twisted transfer matrix \eqref{twist-transfer}. The latter can instead be diagonalized by means of Separation of Variables \cite{Skl90,Skl92,Nic13,NicT15}, and it is worth mentioning that the description of the spectrum in terms of Bethe equations differs from the periodic or diagonal case in what concerns the analytical properties of the $Q$-function \cite{BatBOY95,NicT15}. Due to this difficulty, the method used in \cite{KitMNT16,KitMNT17,KitMNT18} to construct scalar products of separate states cannot be directly generalized to this case, and this is the purpose of the present paper to deal with this problem. From now on, we therefore restrict our study to twists of the form \eqref{form-K}.

\section{Diagonalisation of the transfer matrix by Separation of Variables}
\label{sec-SOV}

In this section, we briefly recall the determination of the spectrum and construction of the eigenstates of the transfer matrix \eqref{twist-transfer} for $\kappa$-twisted antiperiodic boundary conditions, i.e. for a twist of the form \eqref{form-K}. Explicitly, this $\kappa$-twisted antiperiodic transfer matrix takes the form
\begin{equation}\label{anti-transfer}
  \mathcal{T}_K(\lambda)=\kappa^{-1}  B(\lambda)+\kappa\, C(\lambda).
\end{equation}

Let us suppose that $\eta$ is a generic parameter,  i.e. non commensurate with $i\pi$, and that the inhomogeneity parameters are such that
\begin{equation}\label{cond-inh}
   \Xi_i\cap\Xi_j=\emptyset \quad \text{for} \ i\neq j,
   \quad \text{where}\quad 
   \Xi_i=\Big\{\xi_i+k\eta+i k'\pi\mid k\in\{0,-1\}; k'\in\mathbb{Z} \Big\}.
\end{equation}
Under these hypothesis, the diagonalisation of \eqref{anti-transfer} 
was obtained in \cite{Nic13,NicT15} by Sklyanin's version of Separation of Variables (SoV).
There, a basis $\{\ket{\mathbf{h}}, \mathbf{h}\in\{0,1\}^N\}$ of $\mathcal{H}$ and a basis $\{\bra{\mathbf{h}}, \mathbf{h}\in\{0,1\}^N\}$ of $\mathcal{H}^*$ were explicitly constructed such that
\begin{align}
   &\bra{\mathbf{h}}\, D(\lambda )=\prod_{n=1}^N\sinh(\lambda-\xi_n^{(h_n)})\,\bra{\mathbf{h}}
                                                    =d_\mathbf{h}(\lambda)\,\bra{\mathbf{h}},
      \label{D-left}\\
   &\bra{\mathbf{h}}\, C(\lambda )
      =\sum_{a=1}^N \prod_{b\neq a}
        \frac{\sinh \big( \lambda -\xi_b^{(h_b)}\big) }{\sinh \big( \xi_a^{(h_a)}-\xi_b^{(h_b)}\big) }\,
        \delta_{h_a,0}\, d(\xi_a^{(1)})\, \bra{\mathrm{T}_a^+\mathbf{h} },
\label{C-left} \\
   &\bra{\mathbf{h}}\, B(\lambda )
      =-\sum_{a=1}^N \prod_{b\neq a}
         \frac{\sinh \big( \lambda -\xi_b^{(h_b)}\big) }{\sinh \big( \xi_a^{(h_a)}-\xi_b^{(h_b)}\big) }\,
         \delta_{h_a,1}\, a(\xi_a^{(0)})\, \bra{\mathrm{T}_a^{-}\mathbf{h} }.
\label{B-left}
\end{align}
and
\begin{align}
   &D(\lambda )\,\ket{\mathbf{h}} =\prod_{n=1}^N \sinh (\lambda-\xi_n^{(h_n)})\,\ket{\mathbf{h}}
                                                    =d_\mathbf{h}(\lambda)\,\ket{\mathbf{h}}, 
      \label{D-right}\\
   &C(\lambda )\,\ket{\mathbf{h}} 
      =\sum_{a=1}^N \prod_{b\neq a}
        \frac{\sinh \big( \lambda -\xi_b^{(h_b)}\big) }{\sinh \big( \xi_a^{(h_a)}-\xi_b^{(h_b)}\big) }\,
        \delta_{h_a,1}\, d(\xi_a^{(1)})\, \ket{\mathrm{T}_a^-\mathbf{h} },
\label{C-right} \\
   & B(\lambda )\,\ket{\mathbf{h}} 
      =-\sum_{a=1}^N \prod_{b\neq a}
         \frac{\sinh \big( \lambda -\xi_b^{(h_b)}\big) }{\sinh \big( \xi_a^{(h_a)}-\xi_b^{(h_b)}\big) }\,
         \delta_{h_a,0}\, a(\xi_a^{(0)})\, \ket{\mathrm{T}_a^+\mathbf{h} }.
\label{B-right}
\end{align}
Here and in the following, we use the notations:
\begin{align}
  &\xi_n^{(k_n)}=\xi_n-k_n\eta  \qquad \text{for}\quad   k_n\in\{0,1\},
         \label{xi-shift} \\
  &d_\mathbf{h}(\lambda)=\prod_{n=1}^N\sinh(\lambda-\xi_n^{(h_n)}),
         \label{d_h} \\
  &\mathrm{T}_a^\pm(h_1,\ldots,h_N)=(h_1,\ldots,h_a\pm 1,\ldots,h_N).
\end{align}
The action of $A(\lambda)$ on $\ket{\mathbf{h}}$ and on $\bra{\mathbf{h}}$ follows respectively from \eqref{D-right}-\eqref{B-right} and from \eqref{D-left}-\eqref{B-left} by using the quantum determinant relation \eqref{q-det}. 

We have
\begin{equation} \label{sc-hk}
   \moy{\mathbf{h}\, |\, \mathbf{k} } 
   =\frac{\delta _{\mathbf{h},\mathbf{k}}}{V(\xi_1^{(h_1)},\ldots,\xi_N^{(h_N)})},
\end{equation}
where, for any $n$-tuple $(x_1,\ldots,x_n)$, we denote by $V(x_1,\ldots,x_n)$ the following generalized Vandermonde determinant:
\begin{equation}\label{VDM}
      V(x_1,\ldots,x_n)
       = \prod_{\substack{i,j=1 \\ i<j}}^n \sinh(x_j-x_i)
       = \det_{1\le i,j\le n}\left[ \frac{e^{(2j-n-1) x_i}}{2^{j-1}}\right].
\end{equation}

In this SoV basis, the transfer matrix spectrum and eigenstates can be completely characterised in terms of a system of discrete equations in the inhomogeneity parameters of the model \cite{Skl92,Nic13}. Such a characterisation has been more conveniently reformulated in \cite{NicT15} in terms of the solutions of a functional TQ-equation of Baxter's type (in relation with the eigenvalues of the $Q$-operator previously constructed in  \cite{BatBOY95}).

\begin{theorem}\label{th-spectrum} \cite{NicT15}
(Characterisation of the transfer matrix spectrum)

For any fixed $\kappa\in\mathbb{C}\setminus\{0\}$, the $\kappa$-twisted antiperiodic transfer matrix $\mathcal{T}_K(\lambda)$ \eqref{anti-transfer} associated with a twist matrix of the form \eqref{form-K} defines a one-parameter family of commuting operators. All these families are isospectral, i.e. they admit the same set of eigenvalue functions $\Sigma_{\mathcal T}$ independently of the value of $\kappa$.

Under the condition \eqref{cond-inh}, the spectrum of $\mathcal T_K(\lambda)$ is simple and is given by the set of functions $\tau(\lambda)$ such that:
\begin{enumerate}
\item\label{prop1} $\tau(\lambda)$ is an entire function of $\lambda$;
\item\label{prop2}  it satisfies the quasi-periodicity property $\tau(\lambda+i\pi)=(-1)^{N-1}\,\tau(\lambda)$; 
\item\label{prop3} there exists a unique  trigonometric polynomial $Q$ of the form
\begin{equation}\label{Q-form}
    Q(\lambda)=\prod_{j=1}^N\sinh\left(\frac{\lambda-q_j}{2}\right),
    \qquad
\end{equation}
such that $(Q(\xi_j),Q(\xi_j+i\pi))\not=(0,0)$, $\forall j\in\{1,\ldots, N\}$, and that $\tau(\lambda)$ and $Q(\lambda)$ satisfy the functional TQ-relation
\begin{equation}\label{t-Q}
    \tau(\lambda)\, Q(\lambda)= -a(\lambda )\, Q(\lambda -\eta )+d(\lambda)\, Q(\lambda +\eta ).
\end{equation}
\end{enumerate}
Alternatively, $\Sigma_{\mathcal T}$  is given by the set of functions $\tau(\lambda)$ satisfying the properties \ref{prop1}, \ref{prop2} and
\begin{enumerate}
\item[\ref{prop3}'.] \label{prop3'} there exists a unique  trigonometric polynomial $\widehat Q$ of the form
\begin{equation}\label{Qhat-form}
    \widehat{Q}(\lambda)=\prod_{j=1}^N\sinh\left(\frac{\lambda-\hat{q}_j}{2}\right),
    \qquad
\end{equation}
such that $(\widehat Q(\xi_j),\widehat Q(\xi_j+i\pi))\not=(0,0)$, $\forall j\in\{1,\ldots, N\}$, and that $\tau(\lambda)$ and $\widehat Q(\lambda)$ satisfy the functional TQ-relation
\begin{equation}\label{t-Qhat}
    \tau(\lambda)\, \widehat Q(\lambda)= a(\lambda )\, \widehat Q(\lambda -\eta )-d(\lambda)\, \widehat Q(\lambda +\eta ).
\end{equation}
\end{enumerate}
\end{theorem}

Note that, for a given eigenvalue $\tau(\lambda)$ of \eqref{anti-transfer}, the trigonometric polynomials $Q$ satisfying \eqref{t-Q} and $\widehat{Q}$ satisfying \eqref{t-Qhat} are simply related by
\begin{equation}\label{Q-}
    \widehat{Q}(\lambda) =Q(\lambda+i\pi).
\end{equation}
We also recall that these two solutions satisfy the quantum Wronskian relation \cite{NicT15},
\begin{equation}\label{WQ-d}
    \frac{1}{2}\left[
     Q(\lambda)\, \widehat Q(\lambda-\eta)
     +\widehat Q(\lambda)\, Q(\lambda-\eta) \right]=
     \pm \Big(\frac{i}{2}\Big)^N d(\lambda),
\end{equation}
which implies notably the following sum rule for the roots $q_1,\ldots,q_N$:
\begin{equation}\label{sum-rule}
    \sum_{j=1}^N q_j=\sum_{n=1}^N\left(\xi_n-\frac\eta 2\right)+ik\pi,
    \qquad
    k\in\mathbb{Z}.
\end{equation}

As usual, the condition \ref{prop1} on the analyticity of $\tau(\lambda)$ in Theorem~\ref{th-spectrum} can equivalently\footnote{if of course all the roots are pairwise distinct.} be rewritten in terms of the following system of Bethe equations for the roots $q_1,\ldots,q_N$ of $Q(\lambda)$ satisfying \eqref{t-Q}:
\begin{equation}\label{eq-Bethe}
   a(q_j)\, Q(q_j-\eta)=d(q_j)\, Q(q_j+\eta), \qquad j=1,\ldots,N,
\end{equation}
or alternatively for the roots $\hat{q}_1,\ldots,\hat{q}_N$ of $\widehat{Q}(\lambda)$ satisfying \eqref{t-Qhat}:
\begin{equation}\label{eq-Bethe-hat}
   a(\hat q_j)\, \widehat Q(\hat q_j-\eta)=d(\hat q_j)\, \widehat Q(\hat q_j+\eta), \qquad j=1,\ldots,N.
\end{equation}
%
These systems of equations can be conveniently rewritten (at least when none of the roots coincide with an inhomogeneity parameter) 
respectively as
\begin{alignat}{2}
    &\mathfrak{a}_Q(q_j)=1, \qquad & & j=1,\ldots,N,
    \label{eq-bethe}\\
    &\mathfrak{a}_{\widehat Q}(\hat q_j)=1, \qquad & & j=1,\ldots,N,
    \label{eq-bethe-hat}
\end{alignat}
where, for any function $P$, the function $\mathfrak{a}_P$ is defined as
\begin{equation}\label{def-aP}
   \mathfrak{a}_P(\lambda)= \frac{d(\lambda)}{a(\lambda)}\, \frac{P(\lambda+\eta)}{P(\lambda-\eta)}.
\end{equation}

For a given function $P$, $\epsilon\in\{+,-\}$ and $\kappa\in\mathbb{C}^*$, let us define the separate states\footnote{Similarly as in the XXX case \cite{KitMNT16}, we have used the identity
\begin{align*}
     \prod_{n=1}^N \left(-\frac{a(\xi_n)}{d(\xi_n-\eta)}\right)^{\! h_n}\ V(\xi_1^{(h_1)},\ldots,\xi_N^{(h_N)})
     &=V(\xi_1^{(1-h_1)},\ldots,\xi_N^{(1-h_N)})
     =V(\xi_1+h_1\eta,\ldots,\xi_N+h_N\eta)
\end{align*}
so as to rewrite \eqref{separate-r}. }
\begin{align}
  \bra{P,\kappa,\epsilon } 
  &= \sum_{\mathbf{h}\in\{0,1\}^N}
     \prod_{n=1}^N \left\{ (\epsilon\,\kappa)^{h_n}\, P(\xi_n^{(h_n)}) \right\}\, V(\xi_1^{(h_1)},\ldots,\xi_N^{(h_N)})
    \, \bra{\mathbf{h} },
     \label{separate-l}\\
  &= \prod_{n=1}^N [\epsilon\, \kappa\, P(\xi_n-\eta)]\  {}_\mathsf{n\!}\bra{P,\kappa,\epsilon } ,
  \displaybreak[0]\\
   \ket{ P,\kappa,\epsilon } 
  &= \sum_{\mathbf{h}\in\{0,1\}^N}
     \prod_{n=1}^N \left\{ (-\epsilon\,\kappa)^{-h_n}\, \left(\frac{a(\xi_n)}{d(\xi_n-\eta)}\right)^{\! h_n} P(\xi_n^{(h_n)}) \right\}
     \, V(\xi_1^{(h_1)},\ldots,\xi_N^{(h_N)})
\, \ket{\mathbf{h} }
\nonumber\\
     &=\sum_{\mathbf{h}\in\{0,1\}^N}
     \prod_{n=1}^N \left\{ (\epsilon\,\kappa)^{-h_n}\, P(\xi_n^{(h_n)}) \right\}
     \, V(\xi_1^{(1-h_1)},\ldots,\xi_N^{(1-h_N)})
\, \ket{\mathbf{h} }. \label{separate-r}\\
     &= V(\xi_1,\ldots,\xi_N)\,\prod_{n=1}^N \left[\frac\epsilon\kappa\, P(\xi_n-\eta)\right]\  \ket{ P,\kappa,\epsilon }_\mathsf{n} ,
\end{align}
and their normalized\footnote{This normalization is chosen so as to slightly simplify the expression of the scalar products and form factors.} counterparts:
\begin{align}
   &{}_\mathsf{n\!}\bra{P,\kappa,\epsilon }
   =\sum_{\mathbf{h}\in\{0,1\}^N}
     \prod_{n=1}^N \left(\frac{\epsilon}{\kappa}\, \frac{P(\xi_n)}{P(\xi_n-\eta)}\right)^{1- h_n}\, V(\xi_1^{(h_1)},\ldots,\xi_N^{(h_N)})
    \, \bra{\mathbf{h} },
     \label{separate-l-norm}\\
    & \ket{ P,\kappa,\epsilon }_\mathsf{n}  
    =\sum_{\mathbf{h}\in\{0,1\}^N}
     \prod_{n=1}^N \left( \epsilon\,\kappa\, \frac{P(\xi_n)}{P(\xi_n-\eta)}\right)^{1- h_n}
     \, \frac{V(\xi_1^{(1-h_1)},\ldots,\xi_N^{(1-h_N)})}{V(\xi_1,\ldots,\xi_N)}
\, \ket{\mathbf{h} }. \label{separate-r-norm}
\end{align}

\begin{theorem}
(Characterisation of the transfer matrix eigenstates)

Under the hypothesis of Theorem~\ref{th-spectrum}, for a given $\kappa\in\mathbb{C}\setminus\{0\}$, the one-dimensional left and right eigenspaces of $\mathcal T_K(\lambda)$ \eqref{anti-transfer} associated with the eigenvalue $\tau(\lambda)\in\Sigma_{\mathcal T}$ are respectively spanned by the separate states ${}_\mathsf{n\!}\bra{Q,\kappa,+}={}_\mathsf{n\!}\bra{\widehat{Q},\kappa,-}$ and $\ket{Q,\kappa,+}_\mathsf{n}=\ket{\widehat{Q},\kappa,-}_\mathsf{n}$ where $Q$ of the form \eqref{Q-form} satisfies \eqref{t-Q}, and where  $\widehat{Q}$ of the form \eqref{Qhat-form} satisfies \eqref{t-Qhat}.
\end{theorem}

\begin{rem}
The equality between the two normalized versions of the separate states  comes from the relations
\begin{equation}\label{relations-xi}
   \frac{Q(\xi_n-\eta)}{Q(\xi_n)}
   =- \frac{\widehat Q(\xi_n-\eta)}{\widehat Q(\xi_n)},
   \qquad 
   n=1,\ldots,N,
\end{equation}
which is a consequence of \eqref{t-Q} and \eqref{t-Qhat}. These relations also imply that the eigenstates defined as normalized separate states of the form \eqref{separate-l-norm} or \eqref{separate-r-norm} are well-defined since, for each $n\in\{1,\ldots,N\}$, at least one of the two ratios in \eqref{relations-xi} is well-defined. The ratio \eqref{relations-xi} is moreover finite and non-zero since it is equal to $-\tau(\xi_n)/a(\xi_n)=d(\xi_n-\eta)/\tau(\xi_n-\eta)$.
\end{rem}

\begin{rem}
If $\tau(\lambda)$ is an eigenvalue of the $\kappa$-twisted antiperiodic transfer matrix $\mathcal T_K(\lambda)$ with left and right eigenvectors ${}_\mathsf{n\!}\bra{Q,\kappa,+}$ and $\ket{Q,\kappa,+}_\mathsf{n}$, then $-\tau(\lambda)$ is an eigenvalue of $\mathcal T_K(\lambda)$ with left and right eigenvectors ${}_\mathsf{n\!}\bra{Q,\kappa,-}$ and $\ket{Q,\kappa,-}_\mathsf{n}$, and conversely.
\end{rem}

\section{Statement of the main results}
\label{sec-results}

In this section, we present the main results of our article concerning determinant representations of scalar products and form factors. The computations leading to of these results will be exposed in the next sections.

\subsection{The scalar product of two separate states}
\label{subsec-results-SP}

For two given functions $P$ and $Q$, $\epsilon,\epsilon'\in\{+,-\}$ and $\kappa,\kappa'\in\mathbb{C}^*$, the scalar product of the separate states ${}_\mathsf{n\!}\bra{P,\kappa,\epsilon}$ and $\ket{Q,\kappa',\epsilon'}_\mathsf{n}$ built as in  \eqref{separate-l-norm} and \eqref{separate-r-norm} depends only on the product function $(PQ)$ and on the combination $\epsilon\epsilon' \kappa'/\kappa$. As usual for separate states constructed within the SoV approach, this scalar product can be written as  a single determinant of a sum of two dressed generalized Vandermonde matrices as
\begin{align}
   &\mathbf{S}(PQ,\epsilon\epsilon' \kappa'/\kappa)
   ={}_\mathsf{n\!}\moy{P,\kappa,\epsilon\, |\, Q,\kappa',\epsilon'}_\mathsf{n}
   \nonumber\\
   &\qquad
   =\sum_\mathbf{h} \prod_{n=1}^N  \left[ \epsilon \epsilon'\,\frac{\kappa'}{\kappa}\,\frac{(PQ)(\xi_n)}{(PQ)(\xi_n-\eta)}\right]^{1- h_n}
     \, \frac{V(\xi_1^{(1-h_1)},\ldots,\xi_N^{(1-h_N)})}{V(\xi_1,\ldots,\xi_N)}
     \nonumber\\
    &\qquad
    =\frac{\det_{1\le i,j\le N}\left[e^{(2j-N-1) \xi_i}+ \epsilon \epsilon'\,\frac{\kappa'}{\kappa}\,\frac{(PQ)(\xi_i)}{(PQ)(\xi_i-\eta)}\,e^{(2j-N-1) (\xi_i-\eta)}\right]}{\det_{1\le i,j\le N}\left[e^{(2j-N-1) \xi_i}\right]}.
    \label{SP-det-VDM}
\end{align}
As usual in the SoV approach, this determinant representation depends in a very intricate way from the inhomogeneity parameters and is therefore not convenient for the consideration of the physical model.

Let us now suppose in addition that the functions $P$ and $Q$ are trigonometric polynomials of the form
\begin{equation}\label{P-Q-form}
    P(\lambda)=\prod_{j=1}^N\!\sinh\!\left(\frac{\lambda-p_j}{2}\right),
    \qquad
    Q(\lambda)=\prod_{j=1}^N\!\sinh\!\left(\frac{\lambda-q_j}{2}\right),
\end{equation}
where $p_1,\ldots,p_N$, $q_1,\ldots q_N$ are pairwise distinct and
\begin{equation}
    p_1,\ldots,p_N, q_1,\ldots q_N\in\mathbb{C}\setminus \bigcup\limits_{i=1}^N\Xi_i.
    \label{roots-PQ}
\end{equation}

A first result is that quantities of the form \eqref{SP-det-VDM} can be expressed  into some generalized version of the so-called  Izergin determinant \cite{Ize87}, however intertwined in this case by some non-trivial function. Such generalized Izergin determinant representation can then be transformed into some alternative determinant in which rows and columns are labelled by the roots of the function $P$ and $Q$ \eqref{P-Q-form}. More precisely, one has the following representations for the scalar product \eqref{SP-det-VDM} of two separate states:

\begin{prop}\label{prop-SP-1}
The scalar product \eqref{SP-det-VDM} of two separate states  \eqref{separate-l-norm} and \eqref{separate-r-norm} built from $P$ and $Q$ of the form \eqref{P-Q-form}-\eqref{roots-PQ} can be expressed as
\begin{equation}\label{SP-Ize-1}
   \mathbf{S}(PQ,\alpha)
   =\frac{\det_N\left[ \frac{1}{\sinh(\xi_i-p_k)} +\frac{\alpha\,\tilde f_{P,Q}(\xi_i)}{\sinh(\xi_i-p_k-\eta)}\right]}
               {\det_N\left[ \frac{1}{\sinh(\xi_i-p_k)}\right]},
\end{equation}
in which we have set $\alpha=\epsilon\epsilon' \kappa'/\kappa$ and
\begin{equation}\label{tilde-fPQ}
   \tilde f_{P,Q}(u)=\frac{P(u-\eta+i\pi)}{P(u+i\pi)}\frac{Q(u)}{Q(u-\eta)}.
\end{equation}
If moreover $P$ and $Q$ are such that the following condition is satisfied:
\begin{equation}\label{cond-PQ}
   \frac{(PQ)(\xi_n-\eta)}{(PQ)(\xi_n)}=\frac{(PQ)(\xi_n-\eta+i\pi)}{(PQ)(\xi_n+i\pi)},
   \qquad
   \forall\, n\in\{1,\ldots,N\},
\end{equation}
then \eqref{SP-Ize-1} can be rewritten as
\begin{align}\label{SP-demi-Slav}
    \mathbf{S}(PQ,\alpha)
   &=
   \frac{\det_{N}\! \left[ \mathcal{S}^{(\alpha)}(\mathbf{q}\,|\, \mathbf{p})\right]}{\det_{1\le j,k\le N}\big[\coth(\frac{p_k-q_j-\eta}{2})\big]},
   \nonumber\\
   &=
   \frac{\prod_{i,j=1}^N\sinh(\frac{p_i-q_j-\eta}{2})\, \det_{N}\! \left[ \mathcal{S}^{(\alpha)}(\mathbf{q}\,|\, \mathbf{p})\right]}
   {\cosh\! \Big(\frac{\sum_\ell(p_\ell-q_\ell)-N\eta}{2}\Big)\, 
    \prod_{i<j} \!\Big[\sin(\frac{p_i-p_j}{2}) \sinh(\frac{q_j-q_i}{2})\Big]}, \!\!
\end{align}
in which we have defined, for any parameter $\alpha$,
\begin{align}\label{S_PQ-SP}
   \left[\mathcal{S}^{(\alpha)}(\mathbf{q}\,|\, \mathbf{p})\right]_{j,k}
   &\equiv
   \mathsf{S}_{Q,P}^{(\alpha)}(q_j,p_k)
   \nonumber\\
   &=\coth\!\left(\frac{p_k-q_j-\eta}{2}\right)    +\alpha\,\mathfrak{a}_Q(p_k)\, \coth\!\left(\frac{p_k-q_j}{2}\right)
     \nonumber\\
   &\hspace{2cm}
   -2 \alpha \frac{d(p_k)}{a(q_j)}\frac{Q(q_j+\eta)\, P(q_j+i\pi)}{Q(p_k-\eta)\, P(p_k+i\pi)}\frac{1}{\sinh(p_k-q_j)},
\end{align}
with $\mathfrak{a}_Q$ defined as in \eqref{def-aP}.
\end{prop}

The proof of this result is given in Section~\ref{sec-proof-SP}.

\begin{rem}
The condition \eqref{cond-PQ}  is obviously satisfied if, for all $n\in\{1,\ldots,N\}$,
\begin{equation}\label{P-Qprop}
    \frac{P(\xi_n-\eta)}{P(\xi_n)}=-\frac{P(\xi_n-\eta+i\pi)}{P(\xi_n+i\pi)},
    \qquad
    \frac{Q(\xi_n-\eta)}{Q(\xi_n)}=-\frac{Q(\xi_n-\eta+i\pi)}{Q(\xi_n+i\pi)}  .
\end{equation}
Therefore, it is obviously satisfied in the case of transfer matrix eigenstates due to \eqref{relations-xi} (or due to the property \ref{prop2} of Theorem~\ref{th-spectrum} for the functions $ \tau_P(\lambda)$ and $\tau_Q(\lambda)$ built from $P$ and $Q$ as in \eqref{t-Q} or \eqref{t-Qhat}).
Hence, the class of states built from trigonometric polynomials of the form \eqref{P-Q-form}-\eqref{roots-PQ} satisfying \eqref{P-Qprop} contains in particular the class of all transfer matrix eigenstates.
\end{rem}

As already mentioned,  \eqref{SP-Ize-1} is a generalization, in its form, of the formula obtained in \cite{Ize87} for the partition function of the six-vertex model with domain-wall boundary conditions. A crucial difference is however the presence of the non-trivial function $ \tilde f_{P,Q}$ \eqref{tilde-fPQ}.
In its turn, the determinant  in \eqref{SP-demi-Slav}, with rows  and columns labelled by the roots of $Q$ and $P$, is reminiscent  in its form from its analog in the periodic case obtained by Bethe Ansatz \cite{Sla89}, but its matrix elements \eqref{S_PQ-SP} are less symmetric as those of \cite{Sla89}: they seem to contain only parts of the terms that one could expect from a direct generalization of the determinant of \cite{Sla89}. In fact, \eqref{SP-demi-Slav} appears closer to the "square root" of some analog of the determinant of \cite{Sla89}. More precisely, one can show the following property:

\begin{prop}\label{prop-SP-prod}
The scalar products \eqref{SP-det-VDM} of two separate states  \eqref{separate-l-norm} and \eqref{separate-r-norm} built from $P$ and $Q$ of the form \eqref{P-Q-form}-\eqref{roots-PQ} satisfying \eqref{P-Qprop} verify the following property for any parameters $\alpha$ and $\beta$:
\begin{equation}\label{Slav-prod}
    \mathbf{S}(PQ,\alpha) \, \mathbf{S}(PQ,\beta)
    =(-1)^N \frac{e^{\sum_i (\xi_i-p_i)} }{2^{N(N-1)}}  \prod_{i=1}^N\frac{Q(\xi_i)}{Q(\xi_i-\eta)}\,
        \frac{ \det_{N}\left[\bar{\mathcal{S}}^{(\alpha,\beta)}(\mathbf{q}\,|\, \mathbf{p})\right]}{\det_{1\le i,k\le N}\Big[\frac{1}{\sinh(\frac{p_k-q_i-\eta}{2})}\Big]}
\end{equation}
with
\begin{multline}\label{mat-Slav-prod}
    \left[\bar{\mathcal{S}}^{(\alpha,\beta)}(\mathbf{q}\,|\, \mathbf{p})\right]_{j,k}
    =
    \bar{\mathcal{S}}^{(\alpha,\beta)}_{P,Q}(q_j,p_k) 
    \\
    =\left[\frac{\alpha\, e^{-\frac\eta 2}}{\sinh(\frac{p_k-q_j-\eta}{2})} - \frac{1}{\sinh(\frac{p_k-q_j}{2})}\right]
     + \beta e^{-\frac\eta 2}\, \mathfrak{a}_Q(p_k)\left[ \frac{\alpha\, e^{-\frac\eta 2}}{\sinh(\frac{p_k-q_j}{2})} -\frac{1}{\sinh(\frac{p_k-q_j+\eta}{2})}\right]
     \\
    +2\frac{d(p_k)}{d(q_j)}\frac{Q(q_j-\eta)P(q_j+i\pi)}{Q(p_k-\eta)P(p_k+i\pi)}\left(1-\alpha\beta\, e^{-\eta}\, \mathfrak{a}_Q(q_j)\right)
   \frac{e^{\frac{p_k-q_j}{2}}}{\sinh(p_k-q_j)}.
\end{multline}
\end{prop}

In particular, when $\alpha\beta\, e^{-\eta}=1$ and when the roots of $Q$ satisfy in addition the system of Bethe equations \eqref{eq-bethe}, the last line of \eqref{mat-Slav-prod} cancels, and \eqref{mat-Slav-prod} appears as quite similar, in its form, of the determinant of \cite{Sla89}.

\medskip

As an alternative to the representation \eqref{SP-demi-Slav}-\eqref{S_PQ-SP}, we would also like to present some formulas which directly involve the transfer matrix eigenvalues themselves in the case of two twisted eigenstates (not necesserally with the same twist).

\begin{prop}\label{th-SP-tau}
For two arbitrary parameters $\kappa,\kappa'\in\mathbb{C}^*$ and for $\epsilon,\epsilon'\in\{+,-\}$, let ${}_\mathsf{n\!}\bra{P,\kappa,\epsilon}$ be an eigenstate of the $K$-twisted transfer matrix $\mathcal{T}_K(\lambda)$ \eqref{twist-transfer} with $K=\mathrm{diag}(\kappa, \kappa^{-1})\cdot \sigma^x$, with eigenvalue $\epsilon\tau_P(\lambda)$, and let $\ket{ Q,\kappa',\epsilon'}_\mathsf{n}$ be an eigenstate of the $K'$-twisted transfer matrix $\mathcal{T}_K'(\lambda)$ with $K=\mathrm{diag}(\kappa', {\kappa'}^{-1})\cdot \sigma^x$, with eigenvalue $\epsilon'\tau_Q(\lambda)$.
Then the overlap \eqref{SP-det-VDM} between these two eigenstates can be expressed as:
\begin{align}\label{Ize-tau}
  \mathbf{S}(PQ,\epsilon\epsilon' \kappa'/\kappa)
  &=\frac{\det_N\left[ \frac{\tau_Q(\xi_i)}{\sinh(\xi_i-p_k)} -\frac{\kappa'}{\kappa}\frac{\tau_P(\xi_i)}{\sinh(\xi_i-p_k-\eta)}\right]}
               {\det_N\left[ \frac{\tau_Q(\xi_i)}{\sinh(\xi_i-p_k)}\right]},
\end{align}
where $p_1,\ldots,p_N$ are the roots of $P$ \eqref{P-Q-form}.
It can be rewritten as
\begin{equation}\label{SP-tau}
    \mathbf{S}(PQ,\epsilon\epsilon' \kappa'/\kappa)
    =\frac{\prod_{i,j=1}^N\big[\sinh(z_i-\xi_j)\sinh(\xi_j-p_i)\big]\, \det_{N}\left[ \mathcal{S}^{(\kappa'/\kappa)}_{\tau_Q,\tau_P}(\mathbf{z}\, |\,\mathbf{p})\right]}{e^{\sum_\ell\xi_\ell}\, \prod_{j=1}^N\tau_Q(\xi_j)\,\prod_{i<j}\big[\sinh(z_j-z_i)\sinh(p_i-p_j)\big]},
\end{equation}
in which $z_1,\ldots,z_N$ are arbitrary parameters, and 
\begin{align}\label{matS-tau}
  \left[ \mathcal{S}^{(\alpha)}_{\tau_Q,\tau_P}(\mathbf{z}\, |\,\mathbf{p})\right]_{i,k} 
  &=  \mathsf{S}^{(\alpha)}_{\tau_Q,\tau_P}(z_i,p_k)
  \nonumber\\
   &=\frac{\hat\tau_Q(z_i)-\hat\tau_Q(p_k)}{\sinh(z_i-p_k)}-\alpha\frac{\hat\tau_P(z_i)-\hat\tau_P(p_k+\eta)}{\sinh(z_i-p_k-\eta)}.
\end{align}
Here we have used the shortcut notations:
\begin{equation}
 \hat\tau_P(\lambda)=e^\lambda\frac{\tau_P(\lambda)}{d(\lambda)},
 \qquad
  \hat\tau_Q(\lambda)=e^\lambda\frac{\tau_Q(\lambda)}{d(\lambda)}.
\end{equation}
\end{prop}

\begin{rem}
One can of course make particular choices for the parameters $z_1,\ldots,z_N$. One can for instance take them as the roots $q_1,\ldots,q_N$ of $Q$, or alternatively as the roots $p_1,\ldots,p_N$ of $P$.
\end{rem}

This type of representations, directly written in terms of the transfer matrix eigenvalues, may be quite general in the SoV framework, in that it barely relies on the analytical form of the $P$ and $Q$ functions and of the Bethe equations. Indeed, it uses essentially the TQ-equation at the inhomogeneity parameters,
\begin{equation}\label{TQ-xi}
   \epsilon \frac{P(\xi_i-\eta)}{P(\xi_i)}=-\frac{\tau_P(\xi_i)}{a(\xi_i)},
   \qquad
   \epsilon' \frac{Q(\xi_i-\eta)}{Q(\xi_i)}=-\frac{\tau_Q(\xi_i)}{a(\xi_i)},
   \qquad
   i=1,\ldots,N,
\end{equation}
and the properties of the $P$ function at these points, together with the analytical properties of the transfer matrix eigenvalues $\tau_P$ and $\tau_Q$, which are issued from the Yang-Baxter algebra, as for the measures \eqref{VDM}: here $\tau_P$ and $\tau_Q$ are entire $i\pi$-quasi-periodic functions as stated in Theorem~\ref{th-spectrum}. 

Note also that the representation \eqref{Ize-tau} may be interesting in itself, and that a direct consideration of the homogeneous limit in \eqref{Ize-tau} by means of L'Hospital's rule is a possible option at this stage, in that it would lead to an expression involving derivatives of the transfer matrix eigenvalues.

\medskip

Finally, let us make some brief comments about the case $P=Q$. The quantity $\mathbf{S}(Q^2,\alpha)$ can of course be obtained from taking the appropriate limit in the above results. It can also be noticed that, in this case, the function $\tilde f_{Q,Q}$ \eqref{tilde-fPQ} is the constant function equal to $-1$ if $Q$ satisfies \eqref{P-Qprop}. Hence \eqref{SP-Ize-1} reduces to a usual ($\alpha$-twisted) Izergin determinant:
\begin{equation}\label{SP-Ize-P=Q}
   \mathbf{S}(Q^2,\alpha)
   =\frac{\det_N\left[ \frac{1}{\sinh(\xi_i-q_k)} -\frac{\alpha}{\sinh(\xi_i-q_k-\eta)}\right]}
               {\det_N\left[ \frac{1}{\sinh(\xi_i-q_k)}\right]}.
\end{equation}
For this particular case, the same type of transformations as in \cite{KitMNT16,KitMNT17,KitMNT18} can be applied, leading to some kind "Slavnov type" determinant. However, the rows of this resulting determinant are labelled by only half (or a subset) of the roots of $Q$, whereas the columns are labelled by the other half (or the complementary subset). 
Moreover, such a determinant is in that case not expressed in terms of the quantity $\mathfrak{a}_Q$ involved in the Bethe equations, contrary to \eqref{S_PQ-SP}. 
Nevertheless, it should be noticed that \eqref{SP-Ize-P=Q} also allows for alternative types of compact determinant representations. For instance, on can easily transform \eqref{SP-Ize-P=Q} as
\begin{equation}\label{SP-alt-P=Q}
   \mathbf{S}(Q^2,\alpha)
   =\det_{1\le j,k\le N}\left[ \delta_{j,k} -\frac{d(q_k)}{a(q_k)}\,\frac{\prod_{\ell=1}^N\sinh(q_k-q_\ell+\eta)}{\prod_{\ell\not= k}\sinh(q_k-q_\ell)}\, \frac{\alpha}{\sinh(q_k-q_j+\eta)}\right].
\end{equation}
%

\subsection{Form factors of local spin operators}

We now consider the form factors of local spin operators, i.e. the matrix elements of the spin operators acting on a given site $n$ of the chain between two eigenstates of the transfer matrix.
We therefore consider matrix elements of the form
\begin{equation}\label{ff-form}
   {}_\mathsf{n\!}\bra{P,\kappa,\epsilon}\,\sigma_n^\alpha\,\ket{ Q,\kappa,\epsilon'}_\mathsf{n},
   \qquad
   \alpha\in\{+,-,z\},\quad \kappa\in\mathbb{C}\setminus\{0\},\quad \epsilon,\epsilon'\in\{+,-\},
\end{equation}
where ${}_\mathsf{n\!}\bra{P,\kappa,\epsilon}$ and $\ket{ Q,\kappa,\epsilon'}_\mathsf{n}$ are eigenstates of the $K$-twisted transfer matrix \eqref{anti-transfer} with respective eigenvalues $\epsilon\tau_P$ and $\epsilon'\tau_Q$. Due to the equality $\ket{ Q,\kappa,+}_\mathsf{n}=\ket{ \widehat Q,\kappa,-}_\mathsf{n}$, we can without loss of generality restrict our study to the case $\epsilon'=\epsilon$.

\begin{prop}\label{th-ff-sz}
The form factor of the local operator $\sigma_n^z$ in two eigenstates  ${}_\mathsf{n\!}\bra{P,\kappa,\epsilon}$ and $\ket{ Q,\kappa,\epsilon}_\mathsf{n}$ of the $K$-twisted transfer matrix \eqref{anti-transfer} with respective eigenvalues $\epsilon\tau_P$ and $\epsilon\tau_Q$ is given by the following ratio of determinants:
\begin{equation}\label{ff-z-1}
     {}_\mathsf{n\!}\bra{P,\kappa,\epsilon}\,\sigma_n^z\,\ket{ Q,\kappa,\epsilon}_\mathsf{n}
     =  -\prod_{k=1}^n\frac{\tau_P(\xi_k)}{\tau_Q(\xi_k)}\,
   \frac{\det_{N}\left[\mathcal{S}(\mathbf{q}\,|\, \mathbf{p}) 
           -\mathcal{P}^{(z)}(\mathbf{q}\,|\, \mathbf{p},\xi_n)\right]}{\det_{1\le j,k\le N}\big[\coth(\frac{p_k-q_j-\eta}{2})\big]},
\end{equation}
in which $ \mathcal{S}(\mathbf{q}\,|\, \mathbf{p})\equiv  \mathcal{S}^{(1)}(\mathbf{q}\,|\, \mathbf{p})$ is the matrix of the scalar product with elements $ \mathsf{S}_{Q,P}^{(1)}(q_j,p_k)$ \eqref{S_PQ-SP} with $\alpha=1$,
and $\mathcal{P}^{(z)}(\mathbf{q}\,|\, \mathbf{p},\xi_n)$ is a matrix of rank 1 with elements
\begin{align}\label{mat-P-sz}
    \left[ \mathcal{P}^{(z)}(\mathbf{q}\,|\, \mathbf{p},\xi_n)\right]_{j,k}
         &= \frac{P(p_k-\eta)}{Q(p_k-\eta)}\,
     \left[  \frac{Q(\xi_n-\eta)}{P(\xi_n-\eta)}\,\coth\!\left(\frac{\xi_n-q_j-\eta}{2}\right)
     \right.
     \nonumber\\
     &\hspace{2.5cm}\left.
     +\frac{Q(\xi_n-\eta+i\pi)}{P(\xi_n-\eta+i\pi)}\,\coth\!\left(\frac{\xi_n+i\pi-q_j-\eta}{2}\right)
     \right],
\end{align}
for $1\le j,k\le N$. Alternatively, \eqref{ff-z-1} can also be represented as
\begin{multline}\label{ff-z-2}
   _\mathsf{n\!}\bra{P,\kappa,\epsilon}\,\sigma_n^z\,\ket{ Q,\kappa,\epsilon}_\mathsf{n}
   =-\frac{\prod_{i,j=1}^N\big[\sinh(z_i-\xi_j)\sinh(\xi_j-p_i)\big]}{e^{\sum_\ell\xi_\ell}\, \prod_{j=1}^N\tau_Q(\xi_j)\,\prod_{i<j}\big[\sinh(z_j-z_i)\sinh(p_i-p_j)\big]}
   \\
   \times     
   \prod_{k=1}^n\frac{\tau_P(\xi_k)}{\tau_Q(\xi_k)}\
    \det_{N}\left[ \mathcal{S}_{\tau_Q,\tau_P}(\mathbf{z}\, |\,\mathbf{p})
    +\mathcal{P}_{\tau_Q}^{(z)}(\mathbf{z}\, |\,\mathbf{p},\xi_n)\right],
\end{multline}
where $ \mathcal{S}_{\tau_Q,\tau_P}(\mathbf{z}\, |\,\mathbf{p})\equiv \mathcal{S}^{(1)}_{\tau_Q,\tau_P}(\mathbf{z}\, |\,\mathbf{p})$ is the matrix \eqref{matS-tau} and $\mathcal{P}_{\tau_Q}^{(z)}(\mathbf{z}\, |\,\mathbf{p},\xi_n)$ is a matrix of rank 1 with elements
\begin{equation}\label{mat-P-sz-2}
   \left[\mathcal{P}_{\tau_Q}^{(z)}(\mathbf{z}\, |\,\mathbf{p},\xi_n)\right]_{i,\ell}
         =\frac{e^{\xi_n}\,\tau_Q(\xi_n)}{d(p_\ell)\,\sinh(z_i-\xi_n)}\,
      \frac{P(p_\ell-\eta)}{P(\xi_n-\eta)}\frac{P(p_\ell+i\pi)}{P(\xi_n+i\pi)}.
\end{equation}
\end{prop}

\begin{prop}\label{th-ff-s+-}
The form factors of the local operators $\sigma_n^+$ and $\sigma_n^-$ in two eigenstates  ${}_\mathsf{n\!}\bra{P,\kappa,\epsilon}$ and $\ket{ Q,\kappa,\epsilon}_\mathsf{n}$ of the $K$-twisted transfer matrix \eqref{anti-transfer} with respective eigenvalues $\epsilon\tau_P$ and $\epsilon\tau_Q$ coincide, and are given by :
\begin{align}
   {}_\mathsf{n\!}\bra{P,\kappa,\epsilon}\,\sigma_n^-\,\ket{ Q,\kappa,\epsilon}_\mathsf{n}
   &={}_\mathsf{n\!}\bra{P,\kappa,\epsilon}\,\sigma_n^+\,\ket{ Q,\kappa,\epsilon}_\mathsf{n}
   \nonumber\\
   &\hspace{-3cm}=\epsilon\kappa e^{-\sum_j (p_j-\xi_j)} \,  
    \frac{\prod_{k=1}^{n-1}\tau_P(\xi_k)}{\prod_{k=1}^n\tau_Q(\xi_k)}\,
    \nonumber\\
    &\hspace{-1.5cm} \times
            \frac{\det_N\left[ \mathcal{S}^{(e^{-\eta})}(\mathbf{q}\,|\, \mathbf{p}) 
           -\mathcal{P}^{(-)}(\mathbf{q}\,|\, \mathbf{p},\xi_n)\right]
           - \det_N\left[ \mathcal{S}^{(e^{-\eta})}(\mathbf{q}\,|\, \mathbf{p}) \right]}{\det_{1\le j,k\le N}\big[\coth(\frac{p_k-q_j-\eta}{2})\big]},
       \label{ff-s-s+}
\end{align}
in which  $ \mathcal{S}^{(e^{-\eta})}(\mathbf{q}\,|\, \mathbf{p})$ is the matrix of the scalar product \eqref{S_PQ-SP}, and  $\mathcal{P}^{(-)}(\mathbf{q}\,|\, \mathbf{p},\xi_n)$ is a matrix of rank 1 with elements
\begin{multline}\label{mat-P-s-}
    \left[ \mathcal{P}^{(-)}(\mathbf{q}\,|\, \mathbf{p},\xi_n)\right]_{j,k}
         = \frac{e^{-\xi_n+p_k}\,a(\xi_n)\, d(p_k)}{(-2i)^N\,Q(p_k-\eta)\, P(p_k+i\pi)}
     \left[  \frac{Q(\xi_n-\eta)}{P(\xi_n)}\,\coth\!\left(\frac{\xi_n-q_j-\eta}{2}\right)
     \right.
     \\
     \left.
     -\frac{Q(\xi_n-\eta+i\pi)}{P(\xi_n+i\pi)}\,\coth\!\left(\frac{\xi_n-q_j-\eta+i\pi}{2}\right)
     \right],
\end{multline}
for $1\le j,k\le N$.
Alternatively, \eqref{ff-s-s+} can also be represented as
\begin{multline}\label{ff-s-s+2}
   _\mathsf{n\!}\bra{P,\kappa,\epsilon}\,\sigma_n^\pm\,\ket{ Q,\kappa,\epsilon}_\mathsf{n}
   =\epsilon\kappa   
    \frac{e^{-\sum_j p_j} \,\prod_{i,j=1}^N\big[\sinh(z_i-\xi_j)\sinh(\xi_j-p_i)\big]}{ \prod_{j=1}^N\tau_Q(\xi_j)\,\prod_{i<j}\big[\sinh(z_j-z_i)\sinh(p_i-p_j)\big]}
   \\
   \times        
    \frac{\prod_{k=1}^{n-1}\tau_P(\xi_k)}{\prod_{k=1}^n\tau_Q(\xi_k)}\,
   \left( \det_{N}\left[ \mathcal{S}^{(e^{-\eta})}_{\tau_Q,\tau_P}(\mathbf{z}\, |\,\mathbf{p})
    +\mathcal{P}_{\tau_Q}^{(-)}(\mathbf{z}\, |\,\mathbf{p},\xi_n)\right]
    -\det_{N}\left[ \mathcal{S}^{(e^{-\eta})}_{\tau_Q,\tau_P}(\mathbf{z}\, |\,\mathbf{p})\right]\right),
\end{multline}
where $\mathcal{S}^{(e^{-\eta})}_{\tau_Q,\tau_P}(\mathbf{z}\, |\,\mathbf{p})$ is the matrix \eqref{matS-tau} and $\mathcal{P}_{\tau_Q}^{(-)}(\mathbf{z}\, |\,\mathbf{p},\xi_n)$ is a matrix of rank 1 with elements
\begin{equation}\label{mat-P-s--2}
   \left[\mathcal{P}_{\tau_Q}^{(-)}(\mathbf{z}\, |\,\mathbf{p},\xi_n)\right]_{i,k}
         =\frac{e^{p_k}\,a(\xi_n)\,\tau_Q(\xi_n)}{\prod_{\ell=1}^N\sinh(\xi_n-p_\ell)\,\sinh(z_i-\xi_n)}.
\end{equation}
\end{prop}

The computations leading to these results are detailed in Section~\ref{sec-proof-ff}.

\section{Computation of the scalar products}
\label{sec-proof-SP}

In this section, we detail the computations leading to the representations presented in Section~\ref{subsec-results-SP}.

For convenience, we introduce similar notations as in \cite{KitMNT16}: for any set of complex numbers $\{x\}\equiv\{x_1,\dots ,x_M\}$, and a function $f$, we define
\begin{equation}\label{defApm}
\mathcal{A}_{\{ x\}}[ f ] 
      =
      \frac{\det_{1\le i,j\le M}\left[ \frac{e^{(2j-M-1)x_i}}{2^{j-1}}
                -f(x_i)\  \frac{e^{(2j-M-1)(x_i- \eta)}}{2^{j-1}}
                \right] }
               {V(x_1\ldots,x_M) },
\end{equation}
so that the scalar product  \eqref{SP-det-VDM} can be rewritten as
\begin{equation}\label{SP-A}
   \mathbf{S}(PQ,\alpha)
   = \mathcal{A}_{\{\xi_1,\ldots,\xi_N\}}\!\left[-\alpha\,f_{PQ}\right],
 \end{equation}
 where 
 \begin{equation}\label{def-f_PQ}
    f_{PQ}(u)=\frac{(PQ)(u)}{(PQ)(u-\eta)}.
 \end{equation}
 %

\subsection{Proof of Proposition~\ref{prop-SP-1}}
 
The expression \eqref{SP-A} can be easily transformed, similarly as in the XXX case, into some generalized version of the so-called Izergin determinant \cite{Ize87} by means of the following identity. This can be considered as the direct analog of Identity~1 of \cite{KitMNT16}, and can be proven similarly, the main difference being that here we do not a priori obtain an expression which is symmetric in the two considered sets of variables:
%
\begin{identity}
\label{id-Af=If}
For any two sets of arbitrary complex numbers $\{x\}\equiv\{x_1,\dots ,x_L\}$ and $\{z\}=\{z_1,\ldots,z_L\}$ and any function $f$, we have
\begin{equation}
   \mathcal{A}_{\{ x\}}[ f ] 
   =\mathcal{I}_{\{ x\},\{z\}}\left[ f \cdot E_{\{z\}} \right]
\end{equation}
where
\begin{equation}\label{defEpm}
   E_{\{z\}}(u)=\prod_{\ell=1}^L\frac{\sinh(u-z_\ell-\eta)}{\sinh(u-z_\ell)},
\end{equation}
and
\begin{align} \label{def-If}
   \mathcal{I}_{\{ x\},\{z\}}[ f ] 
   &= \frac{\prod_{i,\ell=1}^L\sinh(x_i-z_\ell)}{\prod_{i<j}\sinh(x_j-x_i)\,\sinh(z_i-z_j)}\,
   \det_L\left[ \frac{1}{\sinh(x_i-z_k)}-\frac{f(x_i)}{\sinh(x_i-z_k-\eta)}\right]
   \nonumber\\
   &=\frac{\det_L\left[ \frac{1}{\sinh(x_i-z_k)} -\frac{f(x_i)}{\sinh(x_i-z_k-\eta)}\right]}
               {\det_L\left[ \frac{1}{\sinh(x_i-z_k)}\right]}.
\end{align}
\end{identity}

\begin{proof}
We multiply and divide $\mathcal{A}_{\{ x\}}[ f ]$ with the determinant of the $L\times L$ matrix $\mathcal{C}_{\{z\}}$ with elements $\left(\mathcal{C}_{\{z\}}\right)_{j,k}$, $1\le j,k \le L$, defined as
\begin{equation}
   \prod_{\substack{\ell=1 \\ \ell\not=k}}^L\sinh(\lambda-z_\ell)
   =e^{-(L-1)\lambda}\sum_{j=1}^L \left(\mathcal{C}_{\{z\}}\right)_{j,k}\, \left(\frac{e^{2\lambda}}{2}\right)^{j-1}
   =\sum_{j=1}^L \left( \mathcal{C}_{\{z\}}\right)_{j,k}\, \frac{e^{(2j-L-1)\lambda}}{2^{j-1}},
\end{equation}
and with determinant: $\det_L\mathcal{C}_{\{z\}}=V(z_L,\ldots,z_1)=\prod_{j<k}\sinh(z_j-z_k)$.
\end{proof}

Applying Identity~\ref{id-Af=If} to \eqref{SP-A} with $\{z_1,\ldots,z_L\}$ coinciding with the set $\{p_1,\ldots,p_N\}$ of roots of $P$, we therefore obtain the expression \eqref{SP-Ize-1}:
\begin{align}\label{SP-I}
   \mathbf{S}(PQ,\alpha)
   &= \mathcal{I}_{\{\xi_1,\ldots,\xi_N\},\{p_1,\ldots,p_N\}}\!\left[-\alpha\,\tilde f_{P,Q}\right]=\frac{\det_N \mathcal{M}^{(\alpha)}}{\det_N\mathcal{M}^{(0)}},
 \end{align}
in terms of the function $\tilde f_{P,Q}$ \eqref{tilde-fPQ}, where we have set
\begin{align}\label{M-beta}
    &\mathcal{M}^{(\beta)}_{i,k}=\mathcal{M}^{(\beta)}(\xi_i,p_k)
        = \frac{1}{\sinh(\xi_i-p_k)}+\frac{\beta\, \tilde f_{P,Q}(\xi_i)}{\sinh(\xi_i-p_k-\eta)},
        \qquad
        1\le i,k\le N, 
\end{align}
for $\beta=\alpha$ or $\beta=0$.
The difficulty at this stage is that, contrary to what happens in  \cite{KitMNT16} in the XXX case, we do not obtain a symmetric expression in the two sets of variables, and this even if we micmic the computation of \cite{KitMNT16} (i.e. if we extend the expression \eqref{defApm} to a wider set of variables $\{\xi_1,\ldots,\xi_{2N}\}$ --- see Identity~\ref{id-extension-det} --- and use Indentity~\ref{id-Af=If} with $\{z\}\equiv\{p\}\cup\{q\}$). 
This is due to the fact that the functions $P$ and $Q$  have a different periodicity from the functions involved in the construction of the model. Hence, the set of algebraic transformations used in \cite{KitMNT16} cannot be applied here. We therefore propose an alternative way of transforming the determinant:  the idea is here to multiply both determinants in the numerator and denominator of \eqref{def-If} by the determinant of an adequately chosen matrix $\mathcal{X}$. 

A key remark in the choice of this matrix $\mathcal{X}$ is the fact that, although the functions $f_{PQ}$ \eqref{def-f_PQ} and $\tilde f_{P,Q}$ \eqref{tilde-fPQ} are not $i\pi$-periodic, they become so when evaluated at the inhomogeneity parameters $\xi_1,\ldots,\xi_N$ due to \eqref{cond-PQ}\footnote{In a previous attempt (see arXiv:2011.06109v1), we tried to extend the expression \eqref{defApm} to a wider set of variables $\{\xi_1,\ldots,\xi_{2N}\}$ using Identity~\ref{id-extension-det} before transforming it into a generalized Izergin determinant using Identity~\ref{id-Af=If}. We obtained aa expression proportional to
\begin{equation}
   \lim_{\xi_{N+1}\ldots\xi_{2N}\to +\infty}\mathcal{I}_{\{\xi_1,\ldots,\xi_{2N}\},\{q\}\cup\{p\}}[\beta\bar f_{PQ}],
   \qquad
   \bar f_{PQ}(u)=\frac{(PQ)(u-\eta+i\pi)}{(PQ)(u+i\pi)},
\end{equation}
where $\beta$ is a numerical coefficient. The problem with this approach is that we do not have the effective $i\pi$-periodicity of the function $\bar f_{PQ}$ at $\xi_{N+1},\ldots,\xi_{2N}$, the latter being true only at the limit, i.e. 
\begin{equation}\label{lim-f_PQ-xi}
   \bar f_{PQ}(\xi+i\pi)=\bar f_{PQ}(\xi)+O(e^{-\xi}) \qquad \text{when}\quad \xi\to\infty.
\end{equation}
It happens that the corrections to \eqref{lim-f_PQ-xi}, although exponentially small, do contribute to order 1 to formula (5.19) of arXiv:2011.06109v1 which is therefore not correct. Since taking into account these corrections lead to significantly more complicated formulas, we find it more convenient to use directly \eqref{SP-I}.
}.

By choosing $\mathcal{X}$ as in \eqref{mat-X} below, one obtains the following identity, which directly lead to \eqref{SP-demi-Slav}:

\begin{identity}\label{id-I-Sl-SP}
Let $\{\xi_1,\ldots,\xi_{N}\}$ be a set of arbitrary parameters and let $P$ and $Q$ be trigonometric polynomials of the form \eqref{P-Q-form}-\eqref{roots-PQ},
satisfying moreover the condition \eqref{cond-PQ}.
Let $\{p_1,\ldots,p_N\}$ and $\{q_1,\ldots,q_N\}$ denote their respective sets of roots, and let $\tilde f_{P,Q}$ be the function defined in terms of $P$ and $Q$ as \eqref{tilde-fPQ}.
Then, for any arbitrary parameter $\alpha$,
\begin{equation}\label{I-Sl-SP}
   \mathcal{I}_{\{\xi_1,\ldots,\xi_{N}\},\{p_1,\ldots,p_N\}}
   \big[-\alpha\, \tilde f_{P,Q}\big]
   =\frac{\det_{1\le i,k\le N}\left[\mathsf{S}_{Q,P}^{(\alpha)}(q_j,p_k)\right]}{\det_{1\le j,k\le N}\big[\coth(\frac{p_k-q_j-\eta}{2})\big]},
\end{equation}
where $\mathsf{S}_{Q,P}^{(\alpha)}$ is defined as in \eqref{S_PQ-SP}.
\end{identity}

\begin{proof}
Let us multiply both matrices in \eqref{SP-I} by the same $N\times N$ matrix $\mathcal{X}$ of elements:
\begin{align}\label{mat-X}
    \mathcal{X}_{a,b}
     &\equiv \mathcal{X}(q_a,\xi_b)
     \nonumber\\
    &=\frac{1}{\prod_{\substack{\ell=1 \\ \ell\not=b}}^{N}\sinh(\xi_b-\xi_\ell)}
    \left[Q(\xi_b-\eta)\, P(\xi_b+i\pi)\,\coth\left(\frac{\xi_b-q_a-\eta}{2}\right)\right.
  \nonumber\\
  &\hspace{4cm} \left. -Q(\xi_b-\eta+i\pi)\, P(\xi_b)\,\coth\left(\frac{\xi_b-q_a-\eta+i\pi}{2}\right)\right],
\end{align}
for $1\le a,b\le N$.
Then, 
\begin{equation}
    \mathcal{I}_{\{\xi_1,\ldots,\xi_{N}\},\{p_1,\ldots,p_N\}} \big[-\alpha\, \tilde f_{P,Q}\big]
    =\frac{\det_{N}[\mathcal{X}\,\mathcal{M}^{(\alpha)}]}{\det_{N}[\mathcal{X}\,\mathcal{M}^{(0)}]},
\end{equation}
and we have to compute, for $\beta=\alpha,0$, the elements of the product of matrices
\begin{align}
   \big[\mathcal{X}\,\mathcal{M}^{(\beta)}\big]_{j,k}
   &=\sum_{b=1}^{N} \mathcal{X}_{j,b}\cdot \mathcal{M}^{(\beta)}_{b,k}
     =\sum_{b=1}^{N} \mathcal{X}(q_j,\xi_b)\cdot \mathcal{M}^{(\beta)}(\xi_b,p_k)
     \nonumber\\
   &= \sum_{b=1}^{N}\left[ \mathrm{Res}\left(\mathcal{J}^{(\beta)}_{j,k};\xi_b\right)
 +\mathrm{Res}\left(\mathcal{J}^{(\beta)}_{j,k};\xi_b+i\pi\right)\right].
 \label{sum-Mbeta}
\end{align}
Here, for each $(j,k)\in\{1,\ldots,N\}^2$,  we have defined the function
\begin{align}\label{Ibeta}
    \mathcal{J}^{(\beta)}_{j,k}(\lambda)
    &\equiv \mathcal{J}^{(\beta)}(\lambda\, |\, q_j,p_k) 
    \nonumber\\
    &=\frac{\coth\left(\frac{\lambda-q_j-\eta}{2}\right)}{d(\lambda)}
    \left[\frac{Q(\lambda-\eta)\, P(\lambda+i\pi)}{\sinh(\lambda-p_k)}
     +\beta\,\frac{Q(\lambda)\, P(\lambda-\eta+i\pi)}{\sinh(\lambda-p_k-\eta)}\right].
\end{align}
In \eqref{sum-Mbeta}, we have in particular used the fact $\tilde f_{P,Q}(\xi_b+i\pi)=\tilde f_{P,Q}(\xi_b)$ for any $b\in\{1,\ldots,N\}$ due to \eqref{cond-PQ}.
For each $(j,k)\in\{1,\ldots,N\}^2$, or more generally for any value of $q_j$ and $p_k$, the function \eqref{Ibeta} satisfies the properties
\begin{align}
   &\mathcal{J}^{(\beta)}(\lambda+2\pi i \, |\, q_j,p_k)=\mathcal{J}^{(\beta)}(\lambda\, |\, q_j,p_k),
   \label{prop-J1}\\
   &\mathcal{J}^{(\beta)}(\lambda\, |\, q_j,p_k) = O(e^{- |\Re(\lambda)|}) \qquad
   \text{when}\quad \Re(\lambda)\to\pm\infty,
   \label{prop-J2}
\end{align}
so that the integral of this function on a strip of width $2\pi i$ vanishes:
\begin{align}
   0 &=\oint_{2\pi i\text{-strip}} \frac{dz}{2\pi i} \,\mathcal{J}^{(\beta)}_{j,k}(z)
   \nonumber\\
   &= \sum_{b=1}^{N}\left[ \mathrm{Res}\left(\mathcal{J}^{(\beta)}_{j,k};\xi_b\right)
 +\mathrm{Res}\left(\mathcal{J}^{(\beta)}_{j,k};\xi_b+i\pi\right)\right]
 +\sum_{\text{other poles $z_\ell$}} \hspace{-2mm} \mathrm{Res}\left(\mathcal{J}^{(\beta)}_{j,k}; z_\ell\right),
 \label{id-residues}
\end{align}
where the last sum runs over all the poles $z_\ell$ of $\mathcal{J}^{(\beta)}_{j,k}$ within the $2\pi i\text{-strip}$ which are distinct from $\xi_1,\ldots,\xi_{N}$ (modulo $i\pi $).
Hence,
\begin{align}
   \big[\mathcal{X}\,\mathcal{M}^{(\beta)}\big]_{j,k}
   &= -\mathrm{Res}\left(\mathcal{J}^{(\beta)}_{j,k}; q_j+\eta\right)-\mathrm{Res}\left(\mathcal{J}^{(\beta)}_{j,k}; p_k\right)-\mathrm{Res}\left(\mathcal{J}^{(\beta)}_{j,k}; p_k+\eta\right)
   \nonumber\\
   &= -2\beta\,\frac{Q(q_j+\eta)\, P(q_j+i\pi)}{a(q_j)\,\sinh(q_j-p_k)}
   -\frac{Q(p_k-\eta)\, P(p_k+i\pi)}{d(p_k)}\, \coth\left(\frac{p_k-q_j-\eta}{2}\right)
   \nonumber\\
   &\hspace{2cm}
   -\beta\, \frac{Q(p_k+\eta)\, P(p_k+i\pi)}{a(p_k)}\, \coth\left(\frac{p_k-q_j}{2}\right),
   \label{XMbeta}
\end{align}
which leads to the result when factorizing $ -\frac{Q(p_k-\eta)\, P(p_k+i\pi)}{d(p_k)}$ out of each column $k$ of the determinants.
\end{proof}

Identities~\ref{id-Af=If} and \ref{id-I-Sl-SP} therefore lead to Proposition~\ref{prop-SP-1}.

\begin{rem}
Along the same line, one can of course obtain different types of formulas for the scalar product \eqref{SP-demi-Slav} by using a different matrix $\mathcal{X}$. For instance, instead of \eqref{mat-X}, one can consider, for arbitrary $\gamma$, the following slightly modified matrix $\mathcal{X}_\gamma$:
\begin{align}\label{mat-X-gamma}
    \left[ \mathcal{X}_\gamma\right]_{a,b}
     &\equiv \mathcal{X}_\gamma(q_a,\xi_b)
     \nonumber\\
    &=\frac{1}{\prod_{\substack{\ell=1 \\ \ell\not=b}}^{N}\sinh(\xi_b-\xi_\ell)}
    \big[Q(\xi_b-\eta)\, P(\xi_b+i\pi)\,s_\gamma(\xi_b-q_a-\eta)
  \nonumber\\
  &\hspace{4cm}  -Q(\xi_b-\eta+i\pi)\, P(\xi_b)\,s_\gamma(\xi_b-q_a-\eta+i\pi)\big],
\end{align}
where
\begin{equation}
    s_\gamma(u)=\frac{\sinh\!\left(\frac{u+\gamma}{2}\right)}{\sinh\!\left(\frac u 2\right)\, \sinh\!\left(\frac \gamma 2\right)},
\end{equation}
and obtain the following generalization of \eqref{SP-demi-Slav}:
\begin{align}\label{SP-demi-Slav-gamma}
    \mathbf{S}(PQ,\alpha)
   &=
   \frac{\det_{N}\! \left[ \mathcal{S}^{(\alpha)}_\gamma(\mathbf{q}\,|\, \mathbf{p})\right]}{\det_{1\le j,k\le N}\big[s_\gamma(p_k-q_j-\eta)\big]},
\end{align}
with
\begin{align}\label{S_PQ-SP-gamma}
   \left[\mathcal{S}^{(\alpha)}_\gamma(\mathbf{q}\,|\, \mathbf{p})\right]_{j,k}
      &=s_\gamma(p_k-q_j-\eta)    +\alpha\,\mathfrak{a}_Q(p_k)\, s_\gamma(p_k-q_j)
     \nonumber\\
   &\hspace{2cm}
   -2 \alpha \frac{d(p_k)}{a(q_j)}\frac{Q(q_j+\eta)\, P(q_j+i\pi)}{Q(p_k-\eta)\, P(p_k+i\pi)}\frac{1}{\sinh(p_k-q_j)}.
\end{align}

\end{rem}

\subsection{Proof of Proposition~\ref{prop-SP-prod}}

Let us now show Proposition~\ref{prop-SP-prod}. For this, we will compute some intermediate representations for the generalized Izergin determinant $ \mathcal{I}_{\{\xi_1,\ldots,\xi_{N}\},\{p_1,\ldots,p_N\}}[-\alpha\, \tilde f_{P,Q}]$ \eqref{SP-I} in terms of functions of double period.

\begin{identity}\label{Id-Ize-semi-transformed}
Under the hypothesis of Identity~\ref{id-I-Sl-SP}, and supposing in addition that $P$ and $Q$ satisfy \eqref{P-Qprop}, \eqref{SP-I} can be rewritten as
\begin{align}
    \mathbf{S}(PQ,\alpha)
   &=\frac{e^{\sum_i \frac{\xi_i-p_i}{2}}}{2^N}\, \frac{\det_{1\le i,k\le N} {\mathsf M}^{(\alpha)}_{P,Q}(\xi_i,p_k)}{\det_{1\le i,k\le N}\left[ \frac{1}{\sinh(\xi_i-p_k)}\right]},
    \label{SP-Ize-2}\\
   &=\frac{e^{\sum_i \frac{\xi_i-p_i}{2}}}{2^N} \prod_{i=1}^N \frac{Q(\xi_i)}{P(\xi_i)} \prod_{i<j}\frac{\sinh(\frac{p_j-p_i}{2})}{\sinh(\frac{q_j-q_i}{2})}\,
    \frac{\det_{1\le i,k\le N} \widehat{\mathsf M}^{(\alpha)}_{P,Q}(\xi_i,q_k)}{\det_{1\le i,k\le N}\left[ \frac{1}{\sinh(\xi_i-p_k)}\right]},
    \label{SP-Ize-3}
\end{align}
with
\begin{multline}\label{M-alpha1}
   {\mathsf M}^{(\alpha)}_{P,Q}(\xi_i,p_k)
        =  \frac{1}{\sinh(\frac{\xi_i-p_k}{2})} -\frac{i}{\sinh(\frac{\xi_i-p_k+i\pi}{2})}
        \\
        + \alpha\, e^{-\frac \eta 2}\left( \frac{\tilde f_{P,Q}(\xi_i)}{\sinh(\frac{\xi_i-p_k-\eta}{2})}- \frac{i \,\tilde f_{P,Q}(\xi_i+i\pi)}{\sinh(\frac{\xi_i-p_k-\eta+i\pi}{2})} \right) ,
\end{multline}
and
\begin{align}\label{Mhat-alpha2}
   \widehat{\mathsf M}^{(\alpha)}_{P,Q}(\xi_i,q_k)
   &=\frac{1}{\sinh(\frac{\xi_i-q_k}{2})}-\frac{\alpha\, e^{-\frac\eta 2}}{\sinh(\frac{\xi_i-q_k-\eta}{2})}
   \nonumber\\
   &\quad -i  \frac{P(\xi_i)\, Q(\xi_i+i\pi)}{P(\xi_i+i\pi)\, Q(\xi_i)}\left(\frac{1}{\sinh(\frac{\xi_i-q_k+i\pi}{2})}-\frac{\alpha\, e^{-\frac\eta 2}}{\sinh(\frac{\xi_i-q_k-\eta+i\pi}{2})}\right)
   \nonumber\\
   &=\frac{1}{\sinh(\frac{\xi_i-q_k}{2})}-\frac{\alpha\, e^{-\frac\eta 2}}{\sinh(\frac{\xi_i-q_k-\eta}{2})}
   \nonumber\\
   &\quad +i  \frac{P(\xi_i)\, Q(\xi_i+i\pi-\eta)}{P(\xi_i+i\pi)\, Q(\xi_i-\eta)}\left(\frac{1}{\sinh(\frac{\xi_i-q_k+i\pi}{2})}-\frac{\alpha\, e^{-\frac\eta 2}}{\sinh(\frac{\xi_i-q_k-\eta+i\pi}{2})}\right),
\end{align}
\end{identity}

\begin{proof}
One rewrites the expressions \eqref{SP-I} and \eqref{M-beta} by making explicit the fact that the expression \eqref{M-beta} is in fact $i\pi$-quasi-periodic in $\xi_i$. Using for instance the identity
\begin{equation}\label{id-pi-2pi}
   \frac{1}{\sinh u}=\frac{e^{\frac u 2}}{2}\left(\frac{1}{\sinh\frac u 2}-\frac{i}{\sinh\frac{ u+i\pi} 2}\right),
\end{equation}
we can rewrite \eqref{M-beta} as \eqref{SP-Ize-2}.
Multiplying each row by $P(\xi_i)$, we obtain
\begin{align}
    \mathbf{S}(PQ,\alpha) 
    &=\frac{e^{\sum_i \frac{\xi_i-p_i}{2}}}{2^N\prod_{i=1}^N P(\xi_i)} \frac{\det_N \widetilde{\mathcal M}^{(\alpha)}}{\det_N\mathcal{M}^{(0)}}
\end{align}
with
\begin{align}
    &\widetilde{\mathcal M}^{(\alpha)}_{j,k}=P^{(k)}(\xi_i)
    -\alpha\, e^{-\frac\eta 2}\, \frac{Q(\xi_i)}{Q(\xi_i-\eta)}\, P^{(k)}(\xi_i-\eta)
        \nonumber\\
     &\qquad
    -i  \frac{P(\xi_i)}{P(\xi_i+i\pi)}\left( P^{(k)}(\xi_i+i\pi)-\alpha\, e^{-\frac\eta 2}\,\frac{Q(\xi_i+i\pi)}{Q(\xi_i+i\pi-\eta)}\, P^{(k)}(\xi_i+i\pi-\eta)\right),
    \\
    &P^{(k)}(u) = \frac{P(u)}{\sinh(\frac{u-p_k}{2})},
\end{align}
in which we have used \eqref{P-Qprop} for $P$. Factorizing out the matrix $\widetilde{\mathcal C}_{\{p\},\{q\}}$ with elements defined as
\begin{equation}
    P^{(k)}(u)=\sum_{k=1}^N \left(\widetilde{\mathcal C}_{\{p\},\{q\}}\right)_{j,k} Q^{(j)}(u),
    \qquad \text{with}\quad
    Q^{(j)}(u)=\frac{Q(u)}{\sinh(\frac{u-q_j}{2})},
\end{equation}
of determinant $\det_N\widetilde{\mathcal C}_{\{p\},\{q\}}=\prod_{i<j}\frac{\sinh(\frac{p_j-p_i}{2})}{\sinh(\frac{q_j-q_i}{2})}$, and factorizing $Q(\xi_i)$ out of each row,
we obtain \eqref{SP-Ize-3},
in which in the last equality of \eqref{Mhat-alpha2} we have used \eqref{P-Qprop} for $Q$.
\end{proof}

Using the formulas \eqref{SP-Ize-2} and \eqref{SP-Ize-3}, we can now compute $\mathbf{S}(PQ,\alpha) \, \mathbf{S}(PQ,\beta) $:
\begin{align}
   \mathbf{S}(PQ,\alpha) \, \mathbf{S}(PQ,\beta)
   &=\frac{e^{\sum_i (\xi_i-p_i)}}{2^{2N}} \, \prod_{i=1}^N\frac{Q(\xi_i)}{P(\xi_i)}\,   \prod_{i<j}\frac{\sinh(\frac{p_j-p_i}{2})}{\sinh(\frac{q_j-q_i}{2})}
   \nonumber\\
  &\hspace{2cm} \times 
 \frac{\prod_{i\not= j}\sinh(\xi_i-\xi_j)}{\prod_{i=1}^N P(\xi_i+i\pi)\, Q(\xi_i-\eta)}\,
   \frac{\det_N\bar{\mathcal M}^{(\alpha,\beta)}}{(\det_N\mathcal{M}^{(0)})^2},
\end{align}
with
\begin{align}
   \left[\bar{\mathcal M}^{(\alpha,\beta)}\right]_{i,k} 
   &=\sum_{b=1}^N\prod_{\ell\not= b}\frac{1}{\sinh(\xi_b-\xi_\ell)}
   \left[\left(\frac{1}{\sinh(\frac{\xi_b-q_i}{2})}-\frac{\alpha\, e^{-\frac\eta 2}}{\sinh(\frac{\xi_b-q_i-\eta}{2})}\right)\! P(\xi_b+i\pi) Q(\xi_b-\eta)\right.
   \nonumber\\
   &\hspace{-1cm}
   +i \left.\left(\frac{1}{\sinh(\frac{\xi_b-q_i+i\pi}{2})}-\frac{\alpha\, e^{-\frac\eta 2}}{\sinh(\frac{\xi_b-q_i-\eta+i\pi}{2})}\right)\! P(\xi_b) Q(\xi_b-\eta+i\pi)\right]
   \nonumber\\
   &\hspace{-1cm}
   \times \left[  \frac{1}{\sinh(\frac{\xi_b-p_k}{2})}+\frac{\beta\, e^{-\frac \eta 2} \tilde f_{P,Q}(\xi_b)}{\sinh(\frac{\xi_b-p_k-\eta}{2})} 
        -i\left(\frac{1}{\sinh(\frac{\xi_b-p_k+i\pi}{2})}+\frac{\beta\, e^{-\frac \eta 2} \tilde f_{P,Q}(\xi_b+i\pi)}{\sinh(\frac{\xi_b-p_k-\eta+i\pi}{2})}\right) \right]
        \nonumber\\
    &=\sum_{b=1}^N\left[ \mathrm{Res}\left(\bar{\mathcal J}^{(\alpha,\beta)}_{j,k};\xi_b\right)
 +\mathrm{Res}\left(\bar{\mathcal J}^{(\alpha,\beta)}_{j,k};\xi_b+i\pi\right)\right].
\end{align}
Here we have defined
\begin{align}\label{Ialphabeta}
    \bar{\mathcal J}^{(\alpha,\beta)}_{j,k}(\lambda)
    &\equiv \bar{\mathcal J}^{(\alpha,\beta)}(\lambda\, |\, q_j,p_k) 
    \nonumber\\
    &=\frac{1}{d(\lambda)}
    \left[ \frac{1}{\sinh(\frac{\lambda-q_j}{2})}-\frac{\alpha\, e^{-\frac\eta 2}}{\sinh(\frac{\lambda-q_j-\eta}{2})} \right]
    \nonumber\\
    &\quad
    \times
    \left[ P(\lambda+i\pi)Q(\lambda-\eta)\left( \frac{1}{\sinh(\frac{\lambda-p_k}{2})} -\frac{i}{\sinh(\frac{\lambda-p_k+i\pi}{2})}\right)\right.
    \nonumber\\
    &\quad 
    \left.
    +\beta\, e^{-\frac \eta 2}Q(\lambda)P(\lambda-\eta+i\pi) \left(\frac{1}{\sinh(\frac{\lambda-p_k-\eta}{2})} 
     -   \frac{i}{\sinh(\frac{\lambda-p_k-\eta+i\pi}{2})}\right)\right].
\end{align}
The function \eqref{Ialphabeta} satisfies the same properties \eqref{prop-J1}-\eqref{prop-J2} as \eqref{Ibeta}, so that we can also write:
\begin{align}
   \left[\bar{\mathcal M}^{(\alpha,\beta)}\right]_{i,k} 
   &=-\mathrm{Res}\left(\bar{\mathcal J}^{(\alpha,\beta)}_{j,k}; q_j\right)
       -\mathrm{Res}\left(\bar{\mathcal J}^{(\alpha,\beta)}_{j,k}; q_j+\eta\right)
       \nonumber\\
    &\hspace{2cm}
       - \mathrm{Res}\left(\bar{\mathcal J}^{(\alpha,\beta)}_{j,k}; p_k\right)
       - \mathrm{Res}\left(\bar{\mathcal J}^{(\alpha,\beta)}_{j,k}; p_k+\eta\right)
   \nonumber\\
   &=-2\frac{P(q_j+i\pi)Q(q_j-\eta)}{d(q_j)}\left( \frac{1}{\sinh(\frac{q_j-p_k}{2})} -\frac{i}{\sinh(\frac{q_j-p_k+i\pi}{2})}\right)
    \nonumber\\
    &\quad
   +2\alpha\beta\, e^{-\eta }\frac{Q(q_j+\eta)P(q_j+i\pi)}{a(q_j)} \left(\frac{1}{\sinh(\frac{q_j-p_k}{2})} 
     -   \frac{i}{\sinh(\frac{q_j-p_k+i\pi}{2})}\right)
     \nonumber\\
     &\quad
     -2\frac{P(p_k+i\pi)Q(p_k-\eta)}{d(p_k)}  \left[ \frac{1}{\sinh(\frac{p_k-q_j}{2})}-\frac{\alpha\, e^{-\frac\eta 2}}{\sinh(\frac{p_k-q_j-\eta}{2})} \right]
     \nonumber\\
     &\quad
     -2\beta\, e^{-\frac \eta 2}\frac{Q(p_k+\eta)P(p_k+i\pi)}{a(p_k)} \left[ \frac{1}{\sinh(\frac{p_k-q_j+\eta}{2})}-\frac{\alpha\, e^{-\frac\eta 2}}{\sinh(\frac{p_k-q_j}{2})} \right].
\end{align}
Hence we obtain
\begin{multline}
    \mathbf{S}(PQ,\alpha) \, \mathbf{S}(PQ,\beta)
    =\frac{e^{\sum_i (\xi_i-p_i)}}{2^{2N}}  \frac{\prod_{i\not= j}\sinh(\xi_i-\xi_j)}{\prod_{i=1}^N P(\xi_i)\,P(\xi_i+i\pi)}\, \prod_{i=1}^N\frac{Q(\xi_i)}{Q(\xi_i-\eta)}\,
     \\
    \times
     2^N\prod_{k=1}^N\frac{P(p_k+i\pi)Q(p_k-\eta)}{d(p_k)} 
     \prod_{i<j}\frac{\sinh(\frac{p_j-p_i}{2})}{\sinh(\frac{q_j-q_i}{2})}\,
     \\
     \times
     \frac{\prod_{i,k=1}^N\sinh^2(\xi_i-p_k)}{\prod_{i\not= j}\sinh(\xi_i-\xi_j)\sinh(p_j-p_i)}
     \det_{1\le i,k\le N}\bar{\mathcal{S}}^{(\alpha,\beta)}_{P,Q}(q_j,p_k),
\end{multline}
with $\bar{\mathcal{S}}^{(\alpha,\beta)}_{P,Q}(q_j,p_k)$ given by \eqref{mat-Slav-prod}. Simplifying the prefactor, we obtain \eqref{Slav-prod}.

\subsection{Proof of Proposition~\ref{th-SP-tau}}

The expression \eqref{Ize-tau} follows directly from \eqref{SP-Ize-1} and from the fact that
\begin{equation}\label{fPQ-tau}
   \tilde f_{P,Q}(\xi_i)=-\frac{\epsilon\tau_P(\xi_i)}{\epsilon'\tau_Q(\xi_i)},
   \qquad i=1,\ldots,N.
\end{equation}
Note that all the functions involved in the expression \eqref{Ize-tau} are now explicitly $i\pi$-quasi-periodic.
The expression \eqref{SP-tau} can then be obtained by multiplying   the matrix in the numerator of \eqref{Ize-tau}, with elements,
\begin{align}\label{M-tau}
    &\left[\mathcal{M}^{(\beta)}_\tau\right]_{i,k}=\mathcal{M}^{(\beta)}_\tau(\xi_i,p_k)
        = \frac{\tau_Q(\xi_i)}{\sinh(\xi_i-p_k)}-\frac{\beta\, \tau_P(\xi_i)}{\sinh(\xi_i-p_k-\eta)},
        \quad
        1\le i,k\le N, 
\end{align}
with $\beta=\frac{\kappa'}{\kappa}$, by the matrix $\mathcal{X}_\tau\equiv \mathcal{X}_\tau(\mathbf{z}\, |\, {\boldsymbol \xi})$ of elements
\begin{equation}\label{X-tau}
  \left[ \mathcal{X}_\tau(\mathbf{z}\, |\, {\boldsymbol \xi})\right]_{i,j}
  =\frac{1}{\prod_{\ell\not=j}\sinh(\xi_j-\xi_\ell)}\frac{e^{\xi_j}}{\sinh(z_i-\xi_j)}, 
  \quad 1\le i,j\le N,
\end{equation}
and with determinant
\begin{equation}
  \det_N\left[ \mathcal{X}_\tau\right]
  =\frac{e^{\sum_\ell\xi_\ell}}{\prod_{i,j=1}^N\sinh(z_i-\xi_j)}\,
  \prod_{i<j}\frac{\sinh(z_j-z_i)}{\sinh(\xi_j-\xi_i)}.
\end{equation}
The product $\mathcal{X}_\tau\mathcal{M}^{(\beta)}_\tau$ can be computed as previously:
\begin{align}
   \big[\mathcal{X}_\tau\,\mathcal{M}^{(\beta)}_\tau\big]_{i,k}
   &=\sum_{b=1}^{N} \frac{1}{\prod\limits_{\ell\not=b}\sinh(\xi_b-\xi_\ell)}\frac{e^{\xi_b}}{\sinh(z_i-\xi_b)}
   \left[ \frac{\tau_Q(\xi_b)}{\sinh(\xi_b-p_k)}-\frac{\beta\, \tau_P(\xi_b)}{\sinh(\xi_b-p_k-\eta)}\right]
     \nonumber\\
   &= \sum_{b=1}^{N}\mathrm{Res}\left(\mathcal{J}^{(\tau,\beta)}_{i,k};\xi_b\right),
    \label{sum-Mbeta_tau}
\end{align}
with
\begin{align}\label{Ibeta_tau}
    \mathcal{J}^{(\tau,\beta)}_{j,k}(\lambda)
    &\equiv \mathcal{J}^{(\beta)}_\tau(\lambda\, |\, z_j,p_k) 
    \nonumber\\
    &=-\frac{e^\lambda}{d(\lambda)\, \sinh(\lambda-z_j)}
    \left[\frac{\tau_Q(\lambda)}{\sinh(\lambda-p_k)}
     -\frac{\beta\,\tau_P(\lambda)}{\sinh(\lambda-p_k-\eta)}\right].
\end{align}
For each $(j,k)\in\{1,\ldots,N\}^2$, or more generally for any value of $z_j$ and $p_k$, the function \eqref{Ibeta_tau} satisfies the properties
\begin{align}
   &\mathcal{J}^{(\beta)}_\tau(\lambda+i\pi  \, |\, z_j,p_k)=\mathcal{J}_\tau^{(\beta)}(\lambda\, |\, z_j,p_k),
   \label{prop-Jtau1}\\
   &\mathcal{J}^{(\beta)}_\tau(\lambda\, |\, z_j,p_k) = O(e^{- 2|\Re(\lambda)|}) \qquad
   \text{when}\quad \Re(\lambda)\to\pm\infty,
   \label{prop-Jtau2}
\end{align}
so that the integral of this function on a strip of width $i\pi $ vanishes, and
\begin{align}
   \big[\mathcal{X}_\tau\,\mathcal{M}^{(\beta)}_\tau\big]_{i,k}
   &=-\mathrm{Res}\left(\mathcal{J}^{(\tau,\beta)}_{i,k};z_j\right)
   -\mathrm{Res}\left(\mathcal{J}^{(\tau,\beta)}_{i,k};p_k\right)
   -\mathrm{Res}\left(\mathcal{J}^{(\tau,\beta)}_{i,k};p_k+\eta\right)
\end{align}
which leads to \eqref{SP-tau}-\eqref{matS-tau}.
Here we have used the fact that the transfer matrix eigenvalues $\tau_P(\lambda)$ and $\tau_Q(\lambda)$ are entire functions of $\lambda$, that they satisfy the quasi-periodicity given by item~\ref{prop2} of Theorem~\ref{th-spectrum}, and that they behave as $O(e^{(N-1)|\Re(\lambda)|})$ when $\Re(\lambda)\to\pm\infty$.

\begin{rem}
The function $e^\lambda$ in \eqref{Ibeta_tau} (or equivalently the factor $e^{\xi_j}$ is \eqref{X-tau}) is chosen so as to compensate the difference of $i\pi$-periodicity between the eigenvalues functions $\tau_P(\lambda)$ and $\tau_Q(\lambda)$ and the function $d(\lambda)$ in the denominator of \eqref{Ibeta_tau}. One can of course choose any other entire function such that \eqref{prop-Jtau1}-\eqref{prop-Jtau2} remain valid, the result \eqref{SP-tau}-\eqref{matS-tau} being modified accordingly.
\end{rem}

\subsection{Proof of \eqref{SP-alt-P=Q}}

The representation \eqref{SP-alt-P=Q} cab be obtained from \eqref{SP-Ize-P=Q} along the same lines as above, by multiplying both determinants in \eqref{SP-Ize-P=Q} by the determinant of the same matrix $\widetilde{\mathcal X}$ of elements
\begin{equation}
   \widetilde{\mathcal X}_{a,b}=\frac{\prod_{\ell=1}^N\sinh(\xi_b-q_\ell)}{\prod_{\ell\not= b}\sinh(\xi_b-\xi_\ell)}\frac{1}{\sinh(\xi_b-q_a)},
   \qquad 1\le a,b\le N.
\end{equation}
%

\section{Computation of the form factors}
\label{sec-proof-ff}

As usual, we use the solution of the quantum inverse problem \cite{KitMT99,MaiT00,GohK00} to re-express the local spin operators in terms of the elements of the monodromy matrix dressed by a product of  transfer matrices \eqref{anti-transfer}.

Let $E_n^{i,j}$, $i,j\in\{1,2\}$, be an elementary matrix acting on $V_n$, with elements $(E_n^{i,j})_{k,\ell}=\delta_{i,k}\delta_{j,\ell}$. Similarly as what happens in the periodic case \cite{KitMT99,MaiT00,GohK00}, we can show that (see \cite{Nic13,LevNT16,NicPT20})
\begin{equation}
    E_n^{ij}
     =\prod_{k=1}^{n-1}\mathcal{T}_K(\xi_k)\cdot [T_K(\xi_n) ]_{ji}\cdot \prod_{k=1}^n[\mathcal{T}_K(\xi_k)]^{-1},
      \label{inv-pb1}
\end{equation}
or alternatively that
\begin{align}
     E_n^{ij}
          &=\prod_{k=1}^{n}\mathcal{T}_K(\xi_k)\cdot [T_K(\xi_n)^{-1} ]_{ji}\cdot \prod_{k=1}^{n-1}[\mathcal{T}_K(\xi_k)]^{-1}
          \\
          &=-(-1)^{i+j}\prod_{k=1}^{n}\mathcal{T}_K(\xi_k)\cdot 
          \frac{T_K(\xi_n-\eta)_{3-i,3-j}}{a(\xi_n)\, d(\xi_n-\eta)}\cdot 
          \prod_{k=1}^{n-1}[\mathcal{T}_K(\xi_k)]^{-1}.
     \label{inv-pb2}
\end{align}
Here we have defined, for $K$ being of the form \eqref{form-K},
\begin{equation}\label{anti-mon}
   T_K(\lambda)=K\, T(\lambda)=\begin{pmatrix} \kappa\, C(\lambda) & \kappa\, D(\lambda) \\
   \kappa^{-1} A(\lambda) & \kappa^{-1} B(\lambda) \end{pmatrix}.
\end{equation}
Hence
\begin{align}
    {}_\mathsf{n\!}\bra{P,\kappa,\epsilon}\,\sigma_n^-\,\ket{ Q,\kappa,\epsilon'}_\mathsf{n}
    &=\kappa\,\frac{\prod_{k=1}^{n-1}\epsilon\tau_P(\xi_k)}{\prod_{k=1}^n\epsilon'\tau_Q(\xi_k)}\ 
    {}_\mathsf{n\!}\bra{P,\kappa,\epsilon}\,D(\xi_n)\,\ket{ Q,\kappa,\epsilon'}_\mathsf{n}
    \label{ff_s-1}\\
    &=\kappa^{-1}\,\frac{\prod_{k=1}^{n}\epsilon\tau_P(\xi_k)}{\prod_{k=1}^{n-1}\epsilon'\tau_Q(\xi_k)}\,
    \frac{ {}_\mathsf{n\!}\bra{P,\kappa,\epsilon}\,A(\xi_n-\eta)\,\ket{ Q,\kappa,\epsilon'}_\mathsf{n} }{a(\xi_n)\, d(\xi_n-\eta)},
    \label{ff_s-2}
\end{align}
\begin{align}
    {}_\mathsf{n\!}\bra{P,\kappa,\epsilon}\,\sigma_n^+\,\ket{ Q,\kappa,\epsilon'}_\mathsf{n}
    &=\kappa^{-1}\,\frac{\prod_{k=1}^{n-1}\epsilon\tau_P(\xi_k)}{\prod_{k=1}^n\epsilon'\tau_Q(\xi_k)}\ 
    {}_\mathsf{n\!}\bra{P,\kappa,\epsilon}\,A(\xi_n)\,\ket{ Q,\kappa,\epsilon'}_\mathsf{n},
    \label{ff_s+1}\\
   &=\kappa\,\frac{\prod_{k=1}^{n}\epsilon\tau_P(\xi_k)}{\prod_{k=1}^{n-1}\epsilon'\tau_Q(\xi_k)}\,
    \frac{ {}_\mathsf{n\!}\bra{P,\kappa,\epsilon}\,D(\xi_n-\eta)\,\ket{ Q,\kappa,\epsilon'}_\mathsf{n} }{a(\xi_n)\, d(\xi_n-\eta)},
    \label{ff_s+2}
\end{align}
\begin{align}
    {}_\mathsf{n\!}\bra{P,\kappa,\epsilon}\,\sigma_n^z\,\ket{ Q,\kappa,\epsilon'}_\mathsf{n}
    &=\frac{\prod_{k=1}^{n-1}\epsilon\tau_P(\xi_k)}{\prod_{k=1}^n\epsilon'\tau_Q(\xi_k)}\ 
    {}_\mathsf{n\!}\bra{P,\kappa,\epsilon}\,\left[\kappa\, C(\xi_n)-\kappa^{-1}B(\xi_n)\right]\,\ket{ Q,\kappa,\epsilon'}_\mathsf{n},
    \nonumber\\
    &=
    {}_\mathsf{n\!}\moy{P,\kappa,\epsilon\, |\, Q,\kappa,\epsilon'}_\mathsf{n}
       \nonumber\\
    &\hspace{1.3cm}
    -2\kappa^{-1}
    \frac{\prod_{k=1}^{n-1}\epsilon\tau_P(\xi_k)}{\prod_{k=1}^n\epsilon'\tau_Q(\xi_k)}\ 
    {}_\mathsf{n\!}\bra{P,\kappa,\epsilon}\,B(\xi_n)\,\ket{ Q,\kappa,\epsilon'}_\mathsf{n}.
    \label{ff_sz1}
\end{align} 
In \eqref{ff_sz1}, we have in particular used the fact that the scalar product ${}_\mathsf{n\!}\moy{P,\kappa,\epsilon\, |\, Q,\kappa,\epsilon'}_\mathsf{n}$ vanishes if $\epsilon\tau_P\not=\epsilon'\tau_Q$ so as to simplify the first prefactor.

We can now use the action \eqref{D-left}-\eqref{B-right} of the matrix elements on the SoV state to compute their action on the separate states.

\subsection{The $\sigma_n^z$ form factor}

We use the representation \eqref{ff_sz1}.

Let $\mu$ be an arbitrary parameter. Using \eqref{B-left}, we can compute the action of $B(\mu)$ on the state ${}_\mathsf{n\!}\bra{P,\kappa,\epsilon}$ \eqref{separate-l-norm} as
\begin{align}
   {}_\mathsf{n\!}\bra{P,\kappa,\epsilon}\,B(\mu)
   &= -\epsilon\kappa \, \sum_{\ell=1}^N a(\xi_\ell)\, \frac{P(\xi_\ell-\eta)}{P(\xi_\ell)}\,
   \sum_{\mathbf{h}} \delta_{h_\ell,0} 
   \prod_{n=1}^N \left(\frac{\epsilon}{\kappa}\, \frac{P(\xi_n)}{P(\xi_n-\eta)}\right)^{\! 1- h_n}\, 
   \nonumber\\
   &\hspace{2cm}\times
   \prod_{b\neq \ell}
         \frac{\sinh \big( \mu -\xi_b^{(h_b)}\big) }{\sinh \big( \xi_\ell^{(h_\ell)}-\xi_b^{(h_b)}\big) }\,
         V(\xi_1^{(h_1)},\ldots,\xi_N^{(h_N)})\, \bra{\mathbf{h}},
\end{align}
that we can rewrite as a contour integral as in \cite{NicPT20} as
\begin{align}
   {}_\mathsf{n\!}\bra{P,\kappa,\epsilon}\,B(\mu)
    &=-\epsilon\kappa\, 
    \oint_{ \Gamma_\xi} 
    \frac{dz}{4\pi i}\, \frac{a(z)}{\sinh(\mu-z)}\,   
    \frac{P(z-\eta)}{P(z)}\,
     \sum_{\mathbf{h}} \prod_{b=1}^N\frac{\sinh(\mu-\xi_b^{(h_b)})}{\sinh(z-\xi_b^{(h_b)})}
     \nonumber\\
    &\hspace{2cm}\times    
    \prod_{n=1}^N\left(\frac{\epsilon}{\kappa}\, \frac{P(\xi_n)}{P(\xi_n-\eta)}\right)^{\! 1- h_n}\, 
     V(\xi_1^{(h_1)},\ldots,\xi_N^{(h_N)})\, \bra{\mathbf{h}},
              \label{act-B-int1}
\end{align}
where the contour $\Gamma_\xi$ in \eqref{act-B-int1} surrounds counterclockwise the points $\xi_1,\ldots,\xi_N,\xi_1+i\pi,\ldots,\xi_N+i\pi$, and no other poles of the integrand. Here we have used the condition \eqref{P-Qprop} for the function $P$.
Since the integrand in \eqref{act-B-int1} is a periodic function of $z$ of period $2\pi i$, which behaves as $O(e^{-|\Re(z)|})$ when $\Re(z)\to\pm\infty$, the integral of this function around a strip of width $2\pi i$ vanishes. One can therefore evaluate the integral \eqref{act-B-int1} by considering the residues at the poles outside of  $\Gamma_\xi$ modulo $2\pi i$. Hence we can rewrite \eqref{act-B-int1} as
\begin{align}
   {}_\mathsf{n\!}\bra{P,\kappa,\epsilon}\,B(\mu)
    &=-\epsilon\kappa\, 
    \oint_{ \Gamma_{p,\mu}}
    \frac{dz}{4\pi i}\, \frac{a(z)}{\sinh(z-\mu)}\,   
    \frac{P(z-\eta)}{P(z)}\,
     \sum_{\mathbf{h}} \prod_{b=1}^N\frac{\sinh(\mu-\xi_b^{(h_b)})}{\sinh(z-\xi_b^{(h_b)})}
     \nonumber\\
    &\hspace{2cm}\times    
    \prod_{n=1}^N\left(\frac{\epsilon}{\kappa}\, \frac{P(\xi_n)}{P(\xi_n-\eta)}\right)^{\! 1- h_n}\, 
     V(\xi_1^{(h_1)},\ldots,\xi_N^{(h_N)})\, \bra{\mathbf{h}},
     \label{act-B-int2}
\end{align}
where the contour $\Gamma_{p,\mu}$ surrounds counterclockwise the points $p_1,\ldots,p_N,\mu,\mu+i\pi$, and no other poles of the integrand.

The corresponding matrix element can therefore be written as
\begin{multline}
  {}_\mathsf{n\!}\bra{P,\kappa,\epsilon}\,B(\mu)\,\ket{ Q,\kappa',\epsilon'}_\mathsf{n}
   =-\epsilon\kappa\, \oint_{ \Gamma_{p,\mu}}
   \frac{dz}{4\pi i} \, \frac{a(z)}{\sinh(z-\mu)}\,  \frac{P(z-\eta)}{P(z)}
   \\
   \times
   \sum_\mathbf{h} \prod_{n=1}^N 
   \left\{ \left(\epsilon\epsilon'\,\frac{\kappa'}{\kappa}\, \frac{(PQ)(\xi_n)}{(PQ)(\xi_n-\eta)}\right)^{\! 1- h_n}\, 
   \frac{\sinh(\mu-\xi_n^{(h_n)})}{\sinh(z-\xi_n^{(h_n)})} \right\}  \,      
   \frac{V(\xi_1^{(1-h_1)},\ldots,\xi_N^{(1-h_N)})}{V(\xi_1,\ldots,\xi_N)},
\end{multline}
and, using the notation \eqref{defApm}, as
\begin{multline}\label{el-mat-int}
  {}_\mathsf{n\!}\bra{P,\kappa,\epsilon}\,B(\mu)\,\ket{ Q,\kappa',\epsilon'}_\mathsf{n}
   =-    \epsilon\kappa\, 
\oint_{ \Gamma_{p,\mu}}
   \frac{dz}{4\pi i} \, \frac{a(z)}{\sinh(z-\mu)}\,  \frac{P(z-\eta)}{P(z)}
   \\
   \times
   \prod_{n=1}^N 
            \frac{\sinh(\mu-\xi_n+\eta)}{\sinh(z-\xi_n+\eta)} 
            \    
      \mathcal{A}_{\{\xi_1,\ldots,\xi_N\}}\left[- \alpha\, f_{PQ,\mu,z}\right],
\end{multline}
where we have set $\alpha= \epsilon\epsilon'\frac{\kappa'}{\kappa}$ and where the function $ f_{PQ,\mu,z}$ is defined as
\begin{equation}\label{fPQz}
  f_{PQ,\mu,z}(\lambda)=\frac{(PQ)(\lambda)}{(PQ)(\lambda-\eta)}\,
  \frac{\sinh(\lambda-\mu)\,\sinh(\lambda-\eta-z)}{\sinh(\lambda-\eta-\mu)\,\sinh(\lambda-z)}.
\end{equation}
Let us recall that \eqref{el-mat-int} is in fact a shortcut notation for the sum over the corresponding residues. Writing explicitly this sum, we obtain:
\begin{multline}
\label{el-mat-sum}
  {}_\mathsf{n\!}\bra{P,\kappa,\epsilon}\,B(\mu)\,\ket{ Q,\kappa',\epsilon'}_\mathsf{n}
   =-    \epsilon\kappa\, 
   \Bigg\{ \frac{a(\mu)}{2}\left[ \frac{P(\mu-\eta)}{P(\mu)}- \frac{P(\mu-\eta+i\pi)}{P(\mu+i\pi)}\right]
       \mathcal{A}_{\{\xi_1,\ldots,\xi_N\}}\left[- \alpha\, f_{PQ}\right]
       \\
   +\sum_{\ell=1}^N \frac{a(\mu)}{\sinh(p_\ell-\mu)}\, \frac{P(p_\ell-\eta)}{\prod_{n\not=\ell}\sinh(\frac{p_\ell-p_n}{2})}\,
   \mathcal{A}_{\{\xi_1,\ldots,\xi_N\}}\left[- \alpha\, f^{(\ell)}_{P,Q,\mu}\right]
   \Bigg\},
\end{multline}
where $f_{PQ}$ denotes the function \eqref{def-f_PQ} and
\begin{equation}\label{fell}
   f^{(\ell)}_{P,Q,\mu}(\lambda)
   =\frac{(PQ)(\lambda)}{(PQ)(\lambda-\eta)}\,
  \frac{\sinh(\lambda-\mu)\,\sinh(\lambda-\eta-p_\ell)}{\sinh(\lambda-\eta-\mu)\,\sinh(\lambda-p_\ell)}.\end{equation}
As previously, we can write
 \begin{align}
     \mathcal{A}_{\{\xi_1,\ldots,\xi_N\}}\left[-\alpha\,f_{PQ}\right]
     &=\mathcal{I}_{\{\xi_1,\ldots,\xi_{N}\},\{p_1,\ldots,p_N\}}\left[-\alpha\, \tilde f_{P,Q}\right]
    \nonumber \\
    &= \frac{\det_{N}\! \left[\mathcal{S}^{(\alpha)}(\mathbf{q}\, |\, \mathbf{p})\right]}{\det_{1\le j,k\le N}\big[\coth(\frac{p_k-q_j-\eta}{2})\big]}
    \nonumber\\
    &=\frac{\prod_{i,j=1}^N\big[\sinh(z_i-\xi_j)\sinh(\xi_j-p_i)\big]\, \det_{N}\left[ \mathcal{S}^{(\kappa'/\kappa)}_{\tau_Q,\tau_P}(\mathbf{z}\, |\,\mathbf{p})\right]}{e^{\sum_\ell\xi_\ell}\, \prod_{j=1}^N\tau_Q(\xi_j)\,\prod_{i<j}\big[\sinh(z_j-z_i)\sinh(p_i-p_j)\big]},
\end{align}
in terms of the function $\tilde f_{P,Q}$ \eqref{tilde-fPQ}, of the  matrix of the scalar product \eqref{S_PQ-SP} or of the matrix \eqref{matS-tau}.
We can also transform the determinant in the second line of \eqref{el-mat-sum} similarly.
Using Identity~\ref{id-Af=If}, we can rewrite the determinant as:
\begin{align}
   \mathcal{A}_{\{\xi_1,\ldots,\xi_N\}}\left[- \alpha\, f^{(\ell)}_{P,Q,\mu}\right]
     &=\mathcal{I}_{\{\xi_1,\ldots,\xi_{N}\},\{p^{(\ell)}_1,\ldots,p^{(\ell)}_N\}}\left[-\alpha\, \tilde f_{P,Q}\right],
        \nonumber\\
      &=\prod_{i=1}^{N}\frac{\sinh(\xi_i-\mu)}{\sinh(\xi_i-p_\ell)}
      \prod_{i\not=\ell}\frac{\sinh(p_i-p_\ell)}{\sinh(p_i-\mu)}\
      \frac{\det_{N}\mathcal{M}^{(\alpha;\ell)}}{\det_{N}\mathcal{M}^{(0)}},
        \label{Ize-B}
\end{align}
in terms of the function $\tilde f_{P,Q}$ \eqref{tilde-fPQ}, and where we have set
\begin{equation}\label{def-pell}
   p^{(\ell)}_j=\begin{cases} \mu &\text{if } j=\ell,\\
                      p_j &\text{othewise}.
                      \end{cases}
\end{equation}
and
\begin{align}
    &\mathcal{M}^{(\alpha;\ell)}_{i,k}=\mathcal{M}^{(\alpha)}(\xi_i,p_k^{(\ell)}),\qquad
    \mathcal{M}^{(0)}_{i,k}=\mathcal{M}^{(0)}(\xi_i,p_k),
\end{align}
where $\mathcal{M}^{(\beta)}$ is given by \eqref{M-beta}.

We can now proceed as in the proof of Proposition~\ref{prop-SP-1} to transform \eqref{Ize-B}.
Multiplying the numerator and the denominator in \eqref{Ize-B} by the determinant of the same matrix $\mathcal{X}$ \eqref{mat-X} as for the scalar product,
we obtain
\begin{equation}
   \mathcal{A}_{\{\xi_1,\ldots,\xi_N\}}\left[- \alpha\, f^{(\ell)}_{P,Q,\mu}\right]
   =\prod_{i=1}^{N}\frac{\sinh(\xi_i-\mu)}{\sinh(\xi_i-p_\ell)}
      \prod_{i\not=\ell}\frac{\sinh(p_i-p_\ell)}{\sinh(p_i-\mu)}\
      \frac{\det_{N}\left[\mathcal{X}\mathcal{M}^{(\alpha;\ell)}\right]}{\det_{N}\left[\mathcal{X}\mathcal{M}^{(0)}\right]}.
\end{equation}
The matrix elements of $\mathcal{X}\mathcal{M}^{(0)}$ are given by \eqref{XMbeta} for $\beta=0$. The matrix elements $\big[\mathcal{X}\mathcal{M}^{(\alpha;\ell)}\big]_{j,k}$ for $k\not=\ell$ are given by  \eqref{XMbeta} for $\beta=\alpha$. The matrix elements $\big[\mathcal{X}\mathcal{M}^{(\alpha;\ell)}\big]_{j,\ell}$, $1\le j\le N$, are given by
\begin{align}
   \big[\mathcal{X}\,\mathcal{M}^{(\alpha;\ell)}\big]_{j,\ell} 
   &=\sum_{b=1}^{N} \mathcal{X}_{j,b}\cdot \mathcal{M}^{(\alpha;\ell)}_{b,\ell}
     =\sum_{b=1}^{N} \mathcal{X}(q_j,\xi_b)\cdot \mathcal{M}^{(\alpha)}(\xi_b,\mu)
     \nonumber\\
   &= \sum_{b=1}^{N}\left[ \mathrm{Res}\left(\mathcal{J}^{(\alpha;\ell)}_{j,\ell};\xi_b\right)
 +\mathrm{Res}\left(\mathcal{J}^{(\alpha;\ell)}_{j,\ell};\xi_b+i\pi\right)\right],
   \nonumber\\
   &=-\mathrm{Res}\left(\mathcal{J}^{(\alpha;\ell)}_{j,\ell};q_j+\eta\right)-\mathrm{Res}\left(\mathcal{J}^{(\alpha;\ell)}_{j,\ell};\mu\right)-\mathrm{Res}\left(\mathcal{J}^{(\alpha;\ell)}_{j,\ell};\mu+i\pi\right)
   \nonumber\\
   &\hspace{2cm}
   -\mathrm{Res}\left(\mathcal{J}^{(\alpha;\ell)}_{j,\ell};\mu+\eta\right)-\mathrm{Res}\left(\mathcal{J}^{(\alpha;\ell)}_{j,\ell};\mu+\eta+i\pi\right),
 \label{sum-Mbeta}
\end{align}
with $\mathcal{J}^{(\alpha;\ell)}_{j,\ell}\equiv \mathcal{J}^{(\alpha)}(\lambda\, |\, q_j,\mu)$ as given by the expression \eqref{Ibeta}.
Hence,
\begin{multline}\label{XMell}
   \big[\mathcal{X}\,\mathcal{M}^{(\alpha;\ell)}\big]_{j,\ell} 
   =  -2\alpha\,\frac{Q(q_i+\eta)\, P(q_i+i\pi)}{a(q_i)\,\sinh(q_i-\mu)}
   -\frac{Q(\mu-\eta)\, P(\mu+i\pi)}{d(\mu)}\, \coth\left(\frac{\mu-q_i-\eta}{2}\right)
   \\
    +\frac{Q(\mu-\eta+i\pi)\, P(\mu)}{d(\mu)}\, \coth\left(\frac{\mu-q_i-\eta+i\pi}{2}\right)
      -\alpha\, \frac{Q(\mu+\eta)\, P(\mu+i\pi)}{a(\mu)}\, \coth\left(\frac{\mu-q_i}{2}\right)
      \\
       +\alpha\, \frac{Q(\mu+\eta+i\pi)\, P(\mu)}{a(\mu)}\, \coth\left(\frac{\mu-q_i+i\pi}{2}\right) ,
\end{multline}
so that
\begin{align}
      \mathcal{A}_{\{\xi_1,\ldots,\xi_N\}}\left[- \alpha\, f^{(\ell)}_{P,Q,\mu}\right]
   &=\prod_{i=1}^{N}\frac{\sinh(\xi_i-\mu)}{\sinh(\xi_i-p_\ell)}
      \prod_{i\not=\ell}\frac{\sinh(p_i-p_\ell)}{\sinh(p_i-\mu)}\
 \frac{d(p_\ell)}{d(\mu)}\frac{Q(\mu-\eta)\,P(\mu+i\pi)}{Q(p_\ell-\eta)\,P(p_\ell+i\pi)}
 \nonumber\\
 &\hspace{3cm}\times
 \frac{\det_N\mathcal{S}^{(\alpha;\ell)}_{\mu}(\mathbf{q}\, |\, \mathbf{p})}{\det_N\mathcal{S}^{(0)}(\mathbf{q}\, |\, \mathbf{p})}
   \nonumber\\
   &=\sinh(\mu-p_\ell)\, \frac{P'(p_\ell)}{P(\mu)}\, \frac{Q(\mu-\eta)}{Q(p_\ell-\eta)}\, \frac{\det_N\mathcal{S}^{(\alpha;\ell)}_{\mu}(\mathbf{q}\, |\, \mathbf{p})}{\det_N\mathcal{S}^{(0)}(\mathbf{q}\, |\, \mathbf{p})}.
\end{align}
The elements of $\mathcal{S}^{(\alpha;\ell)}_{\mu}(\mathbf{q}\, |\, \mathbf{p})$ are
\begin{align}
  &\left[\mathcal{S}^{(\alpha;\ell)}_{\mu}(\mathbf{q}\, |\, \mathbf{p})\right]_{j,k} = \left[\mathcal{S}^{(\alpha)}(\mathbf{q}\, |\, \mathbf{p})\right]_{j,k}=
   \mathsf{S}_{Q,P}^{(\alpha)}(q_j,p_k)
   \qquad \text{for}\quad k\not=\ell,
   \\
  & \left[ \mathcal{S}^{(\alpha;\ell)}_{\mu}(\mathbf{q}\, |\, \mathbf{p})\right]_{j,\ell} =\mathsf{S}_{Q,P}^{(\alpha)}(q_j,\mu)
    -\frac{Q(\mu-\eta+i\pi)\,P(\mu)}{Q(\mu-\eta)\, P(\mu+i\pi)}\, 
  \nonumber\\
  &\hspace{1cm}
  \times
\left[ \coth\left(\frac{\mu-q_j-\eta+i\pi}{2}\right)+\alpha\, \mathfrak{a}_Q(\mu+i\pi)\, \coth\!\left(\frac{\mu-q_j+i\pi}{2}\right)\right],
\label{Sell-mu}
\end{align}
where $\mathsf{S}^{(\beta)}_{Q,P}(q_j,\mu)$ is given by \eqref{S_PQ-SP} with $p_k$ replaced by $\mu$.
Hence we have
\begin{multline}
\label{el-mat-sum-2}
  {}_\mathsf{n\!}\bra{P,\kappa,\epsilon}\,B(\mu)\,\ket{ Q,\kappa',\epsilon'}_\mathsf{n}
   = \frac{-    \epsilon\kappa\,a(\mu)}{2\det_{1\le j,k\le N}\big[\coth(\frac{p_k-q_j-\eta}{2})\big]}
   \\
   \times
   \Bigg\{ \left[ \frac{P(\mu-\eta)}{P(\mu)}- \frac{P(\mu-\eta+i\pi)}{P(\mu+i\pi)}\right]
   \det_{N}\! \left[\mathcal{S}^{(\epsilon\epsilon'\kappa'/\kappa)}(\mathbf{q}\, |\, \mathbf{p})\right]
       \\
   -\sum_{\ell=1}^N  \frac{P(p_\ell-\eta)}{P(\mu)}\, \frac{Q(\mu-\eta)}{Q(p_\ell-\eta)}\,\det_N\mathcal{S}^{(\epsilon\epsilon'\kappa'/\kappa;\ell)}_{\mu}(\mathbf{q}\, |\, \mathbf{p})
   \Bigg\}.
\end{multline}
Setting $\mu=\xi_n$, using \eqref{P-Qprop} for $P$ and the fact that
\begin{equation}\label{tauPxi}
  \tau_P(\xi_n)=-a(\xi_n)\frac{P(\xi_n-\eta)}{P(\xi_n)},
\end{equation}
we obtain
\begin{multline}
\label{el-mat-sum-3}
  {}_\mathsf{n\!}\bra{P,\kappa,\epsilon}\,B(\xi_n)\,\ket{ Q,\kappa,\epsilon'}_\mathsf{n}
   =   \frac{\epsilon\kappa\, \tau_P(\xi_n)}{\det_{1\le j,k\le N}\big[\coth(\frac{p_k-q_j-\eta}{2})\big]}\, 
     \Bigg\{ 
     \det_{N}\left[\mathcal{S}^{(\epsilon\epsilon')}(\mathbf{q}\, |\, \mathbf{p})\right]
      \\
     -\frac{1}{2} \frac{Q(\xi_n-\eta)}{P(\xi_n-\eta)}
     \sum_{\ell=1}^N \frac{P(p_\ell-\eta)}{Q(p_\ell-\eta)}\, 
     \det_N\left[\mathcal{S}^{(\epsilon\epsilon';\ell)}_{\xi_n}(\mathbf{q}\, |\, \mathbf{p})\right] \Bigg\},
\end{multline}
and we get from \eqref{ff_sz1} and \eqref{SP-demi-Slav}
\begin{multline}
    {}_\mathsf{n\!}\bra{P,\kappa,\epsilon}\,\sigma_n^z\,\ket{ Q,\kappa,\epsilon'}_\mathsf{n}
  =-\frac{1}{\det_{1\le j,k\le N}\big[\coth(\frac{p_k-q_j-\eta}{2})\big]}\,\
     \prod_{k=1}^n\frac{\epsilon\tau_P(\xi_k)}{\epsilon'\tau_Q(\xi_k)}\,
      \\
   \times
   \Bigg\{ 
   \det_N\left[\mathcal{S}^{(\epsilon\epsilon')}(\mathbf{q}\, |\, \mathbf{p})\right]
     - \frac{Q(\xi_n-\eta)}{P(\xi_n-\eta)}
     \sum_{\ell=1}^N \frac{P(p_\ell-\eta)}{Q(p_\ell-\eta)}\, 
      \det_N\left[\mathcal{S}^{(\epsilon\epsilon';\ell)}_{\xi_n}(\mathbf{q}\, |\, \mathbf{p})\right]
      \Bigg\},
\end{multline}
where
\begin{align}
\left[ \mathcal{S}^{(\epsilon\epsilon';\ell)}_{\xi_n}(\mathbf{q}\, |\, \mathbf{p}) \right]_{j,\ell}
  &= \coth\!\left(\frac{\xi_n-q_j-\eta}{2}\right) 
  \nonumber\\
  &\hspace{1cm}
   -\frac{P(\xi_n)\, Q(\xi_n-\eta+i\pi)}{P(\xi_n+i\pi)\, Q(\xi_n-\eta)}\,
    \coth\!\left(\frac{\xi_n-q_j-\eta+i\pi}{2}\right) .
\end{align}
Hence
\begin{equation}
     {}_\mathsf{n\!}\bra{P,\kappa,\epsilon}\,\sigma_n^z\,\ket{ Q,\kappa,\epsilon'}_\mathsf{n}
     =
   -\prod_{k=1}^n\frac{\epsilon\tau_P(\xi_k)}{\epsilon'\tau_Q(\xi_k)}\,
   \frac{\det_{N}\left[\mathcal{S}^{(\epsilon\epsilon')}(\mathbf{q}\,|\, \mathbf{p}) 
           -\mathcal{P}^{(z)}(\mathbf{q}\,|\, \mathbf{p},\xi_n)\right]}{\det_{1\le j,k\le N}\big[\coth(\frac{p_k-q_j-\eta}{2})\big]}
\end{equation}
in which $ \mathcal{S}^{( \epsilon\epsilon')}(\mathbf{q}\,|\, \mathbf{p})$ is the matrix of the scalar product with elements $\mathsf{S}_{Q,P}^{( \epsilon\epsilon')}(q_j,p_k) $, $1\le j,k \le N$, and $\mathcal{P}^{(z)}(\mathbf{q}\,|\, \mathbf{p},\xi_n)$ is a matrix of rank 1 with elements
\begin{align}\label{mat-P-sz-proof}
    \left[ \mathcal{P}^{(z)}(\mathbf{q}\,|\, \mathbf{p},\xi_n)\right]_{j,k}
    &= \frac{P(p_k-\eta)}{Q(p_k-\eta)}\, \frac{Q(\xi_n-\eta)}{P(\xi_n-\eta)}\,
         \left[ \mathcal{S}^{(\epsilon\epsilon';k)}_{\xi_n}(\mathbf{q}\, |\, \mathbf{p}) \right]_{j,k}
         \nonumber\\
     &= \frac{P(p_k-\eta)}{Q(p_k-\eta)}\,
     \left[  \frac{Q(\xi_n-\eta)}{P(\xi_n-\eta)}\, \coth\!\left(\frac{\xi_n-q_j-\eta}{2}\right)
     \right.
     \nonumber\\
     &\hspace{2cm}\left.
     +\frac{Q(\xi_n-\eta+i\pi)}{P(\xi_n-\eta+i\pi)}\,\coth\!\left(\frac{\xi_n-q_j-\eta+i\pi}{2}\right)
     \right],
\end{align}
in which we have used \eqref{P-Qprop}.

Alternatively, one can transform \eqref{Ize-B} in the line of what has been done for Proposition~\ref{th-SP-tau} using \eqref{fPQ-tau} and rewriting $\mathcal{M}^{(\alpha;\ell)}$ as
\begin{align}
  \mathcal{M}^{(\alpha;\ell)}_{i,k}
  &=\frac{1}{\tau_Q(\xi_i)} \left[\mathcal{M}^{(\alpha;\ell)}_\tau\right]_{i,k}
  = \frac{1}{\tau_Q(\xi_i)}\, \mathcal{M}^{(\beta)}_\tau(\xi_i,p_k^{(\ell)}),
\end{align}
in terms of \eqref{M-tau} witth $\beta=\frac{\kappa'}\kappa$.
Multiplying then by $\mathcal{X}_\tau\equiv \mathcal{X}_\tau(\mathbf{z}\, |\, {\boldsymbol \xi})$ \eqref{X-tau}, we obtain
\begin{align}
   \sum_{b=1}^N \left[\mathcal{X}_\tau\right]_{i,b}\,\mathcal{M}^{(\beta)}_\tau(\xi_b,p_k^{(\ell)})
   &=\mathsf{S}^{(\beta)}_{\tau_Q,\tau_P}(z_i,p_k^{(\ell)})
   \label{mat-Stau-pell}
\end{align}
in terms of the expression \eqref{matS-tau}, so that
\begin{multline}
   \mathcal{A}_{\{\xi_1,\ldots,\xi_N\}}\left[- \alpha\, f^{(\ell)}_{P,Q,\mu}\right]
     =\frac{d(\mu)}{d(p_\ell)}
      \prod_{i\not=\ell}\frac{\sinh(p_i-p_\ell)}{\sinh(p_i-\mu)}\
      \\
      \times
      \frac{\prod_{i,j=1}^N\big[\sinh(z_i-\xi_j)\sinh(\xi_j-p_i)\big]\, \det_{N}\left[ \mathcal{S}^{(\kappa'/\kappa)}_{\tau_Q,\tau_P}(\mathbf{z}\, |\,\mathbf{p}^{(\ell)})\right]}{e^{\sum_\ell\xi_\ell}\, \prod_{j=1}^N\tau_Q(\xi_j)\,\prod_{i<j}\big[\sinh(z_j-z_i)\sinh(p_i-p_j)\big]},
\end{multline}
where the elements of the matrix $\mathcal{S}^{(\kappa'/\kappa)}_{\tau_Q,\tau_P}(\mathbf{z}\, |\,\mathbf{p}^{(\ell)})$ are given by \eqref{mat-Stau-pell}.
Hence, combining these expressions with \eqref{el-mat-sum}, we obtain
\begin{multline}
\label{el-mat-sum-tau}
  {}_\mathsf{n\!}\bra{P,\kappa,\epsilon}\,B(\mu)\,\ket{ Q,\kappa',\epsilon'}_\mathsf{n}
   =-   
   \frac{ \epsilon\kappa\, \prod_{i,j=1}^N\big[\sinh(z_i-\xi_j)\sinh(\xi_j-p_i)\big]}{e^{\sum_\ell\xi_\ell}\, \prod_{j=1}^N\tau_Q(\xi_j)\,\prod_{i<j}\big[\sinh(z_j-z_i)\sinh(p_i-p_j)\big]}
   \\
   \times
   \Bigg\{ \frac{a(\mu)}{2}\left[ \frac{P(\mu-\eta)}{P(\mu)}- \frac{P(\mu-\eta+i\pi)}{P(\mu+i\pi)}\right]
    \det_{N}\left[ \mathcal{S}^{(\kappa'/\kappa)}_{\tau_Q,\tau_P}(\mathbf{z}\, |\,\mathbf{p})\right]
           \\
   +\sum_{\ell=1}^N  \frac{a(\mu)}{\sinh(p_\ell-\mu)}\, \frac{P(p_\ell-\eta)}{\prod_{n\not=\ell}\sinh(\frac{p_\ell-p_n}{2})}\, \frac{d(\mu)}{d(p_\ell)}
      \prod_{i\not=\ell}\frac{\sinh(p_i-p_\ell)}{\sinh(p_i-\mu)}\,
    \det_{N}\left[ \mathcal{S}^{(\kappa'/\kappa)}_{\tau_Q,\tau_P}(\mathbf{z}\, |\,\mathbf{p}^{(\ell)})\right]   \Bigg\}.
\end{multline}
Noticing that
\begin{equation}
    d(\mu)\, \mathsf{S}^{(\alpha)}_{\tau_Q,\tau_P}(z_i,\mu) 
    \underset{\mu\to\xi_n}{\longrightarrow} -\frac{e^{\xi_n}\,\tau_Q(\xi_n)}{\sinh(z_i-\xi_n)},
\end{equation}
using \eqref{tauPxi} and combining  with \eqref{ff_sz1}, we obtain the representation \eqref{ff-z-2}.

\subsection{The $\sigma_n^-$ and $\sigma_n^+$  form factors}

Considering that we have used here the SoV basis that diagonalizes the $D$-operator, it is convenient to compute the $\sigma_n^-$ and $\sigma_n^+$  form factors by means of the representations \eqref{ff_s-1} and \eqref{ff_s+2} respectively. 
We also restrict our study to the case $\epsilon=\epsilon'$.

Let therefore compute, for $\mu$ arbitrary parameter, the matrix element ${}_\mathsf{n\!}\bra{P,\kappa,\epsilon}\,D(\mu)\,\ket{ Q,\kappa,\epsilon}_\mathsf{n}$. Using \eqref{D-left} or \eqref{D-right}, we obtain
\begin{multline}
  {}_\mathsf{n\!}\bra{P,\kappa,\epsilon}\,D(\mu)\,\ket{ Q,\kappa,\epsilon}_\mathsf{n}
   =\sum_\mathbf{h} \prod_{n=1}^N  \left\{  \left[ \frac{(PQ)(\xi_n)}{(PQ)(\xi_n-\eta)}\right]^{1- h_n} 
   \sinh(\mu-\xi_n^{(h_n)}) \right\}  \, 
   \\
   \times
   \frac{V(\xi_1^{(1-h_1)},\ldots,\xi_N^{(1-h_N)})}{V(\xi_1,\ldots,\xi_N)},
\end{multline}
which can be expressed as a single determinant as
\begin{align}\label{el-D-A}
  {}_\mathsf{n\!}\bra{P,\kappa,\epsilon}\,D(\mu)\,\ket{ Q,\kappa,\epsilon}_\mathsf{n}
    &=\prod_{n=1}^N\sinh(\mu-\xi_n+\eta)\,    \mathcal{A}_{\{\xi_1,\ldots,\xi_N\}}[- f_{PQ,\mu}],
\end{align}
in which we have used the notation \eqref{defApm} with the function $f_{PQ,\mu}$ defined as
\begin{equation}
   f_{PQ,\mu}(\lambda)=\frac{(PQ)(\lambda)}{(PQ)(\lambda-\eta)}\,
   \frac{\sinh(\lambda-\mu)}{\sinh(\lambda-\mu-\eta)}.
\end{equation}

We now use the following identity to extend the size of the determinant, by introducing, as in \cite{KitMNT16}, some extra variable tending to $+\infty$:
%
\begin{identity}\label{id-extension-det}
Let us suppose that $f(\lambda)$ has a finite limit when $\lambda\to +\infty$: $f(\lambda)\underset{\lambda\to +\infty}{\longrightarrow} f_\infty$. Then
  \begin{equation}\label{lim-xL}
    \lim_{x_L\to +\infty} \mathcal{A}_{\{ x_1,\ldots,x_L\}}[ f ] 
    = \left(1-f_\infty\, e^{-(L-1)\eta}\right)\,  \mathcal{A}_{\{ x_1,\ldots,x_{L-1}\}}[ e^{\eta} f ] .
  \end{equation}
  More generally,
  \begin{align}\label{lim-x}
    \lim_{x_L, \ldots,x_{L-p+1}\to +\infty} \mathcal{A}_{\{ x_1,\ldots,x_L\}}[ f ] 
     &= \prod_{k=1}^p \left(1-f_\infty\, e^{- (L-2k+1)\eta}\right)\, \mathcal{A}_{\{ x_1,\ldots,x_{L-p}\}}[ e^{p\eta} f ].
  \end{align}
\end{identity}

Using Identity~\ref{id-extension-det} and Identity~\ref{id-Af=If}, we can express $\mathcal{A}_{\{\xi_1,\ldots,\xi_N\}}[-f_{PQ,\mu}]$ in \eqref{el-D-A} as:
\begin{align}
   \mathcal{A}_{\{\xi_1,\ldots,\xi_N\}}[-f_{PQ,\mu}]
   &=\frac{1}{2}
   \lim_{\xi_{N+1}\to \infty}
    \mathcal{A}_{\{\xi_1,\ldots,\xi_{N+1}\}}[-e^{-\eta}\, f_{PQ,\mu}]
    \nonumber\\
   &=\frac{1}{2}
\lim_{\xi_{N+1}\to \infty}
   \mathcal{I}_{\{\xi_1,\ldots,\xi_{N+1}\},\{p_1,\ldots,p_N,\mu\}}[-e^{-\eta}\, \tilde f_{P,Q}]
   \label{Ize-D}
\end{align}
in terms of the same function $\tilde{f}_{P,Q}$ as in \eqref{tilde-fPQ}.
We now need to adapt the proof of Identity~\ref{id-I-Sl-SP} to transform the generalized Izergin determinant \eqref{Ize-D}.

The generalized Izergin determinant $ \mathcal{I}_{\{\xi_1,\ldots,\xi_{N+1}\},\{p_1,\ldots,p_N,\mu\}}[-\alpha\,  \tilde f_{P,Q}]$ is given by the following ratio of determinants:
\begin{align}\label{ratio-Imu}
    \mathcal{I}_{\{\xi_1,\ldots,\xi_{N+1}\},\{p_1,\ldots,p_N,\mu\}}[-\alpha\,  \tilde f_{P,Q}]
    &=\frac{\det_{N+1}\mathcal{M}^{(\alpha)}_\mu}{\det_{N+1}\mathcal{M}^{(0)}_\mu},
    \nonumber\\
    &=\frac{\sinh(\xi_{N+1}-\mu)\prod_{j=1}^N\sinh(\xi_{N+1}-p_j)\sinh(\xi_j-\mu)}{\prod_{j=1}^N\sinh(\xi_{N+1}-\xi_j)\sinh(p_j-\mu)}
    \nonumber\\
    &\hspace{2cm}
    \times
    \frac{\det_{N+1}\mathcal{M}^{(\alpha)}_\mu}{\det_{N}\mathcal{M}^{(0)}},
\end{align}
in which, for any given parameter $\beta$, we define the $(N+1)\times(N+1)$ matrix $\mathcal{M}^{(\beta)}_\mu$ as an extension of the matrix $\mathcal{M}^{(\beta)}$ \eqref{M-beta}:
\begin{alignat}{2}
    \left[ \mathcal{M}^{(\beta)}_\mu \right]_{i,k}
    &= \mathcal{M}^{(\beta)}(\xi_i,p_k) 
    =\frac{1}{\sinh(\xi_i-p_k)}+\frac{\beta\, \tilde f_{P,Q}(\xi_i)}{\sinh(\xi_i-p_k-\eta)}
    & &\quad \text{for } k\le N,
    \label{M-mu-1}\\
    &= \mathcal{M}^{(\beta)}(\xi_i,\mu)
    =\frac{1}{\sinh(\xi_i-\mu)}+\frac{\beta\, \tilde f_{P,Q}(\xi_i)}{\sinh(\xi_i-\mu-\eta)}
    \quad & &\quad \text{for } k=N+1. \hspace{-1cm}
    \label{M-mu-2}
\end{alignat}
Note that the extra column \eqref{M-mu-2} has the same form as the $\ell$-th modified column of the matrix $\mathcal{M}^{(\alpha;\ell)}$ in \eqref{Ize-B}.

Let us multiply both determinants in \eqref{ratio-Imu} by $\det_N\mathcal{X}$, where $\mathcal{X}$ is the matrix \eqref{mat-X}.
Then 
\begin{align}
   \frac{\det_{N+1}\mathcal{M}^{(\alpha)}_\mu}{\det_{N}\mathcal{M}^{(0)}}
   =\frac{\det_{N+1}\widetilde{S}^{(\alpha)}_\mu}{\det_N[\mathcal{X}\mathcal{M}^{(0)}]}
\end{align}
with
\begin{align}
  &\left[ \widetilde{S}^{(\alpha)}_\mu\right]_{j,k}
  = \left[\mathcal{X}\mathcal{M}^{(\alpha)}\right]_{j,k} \qquad \text{if } j,k\le N,
  \\
  &\left[ \widetilde{S}^{(\alpha)}_\mu\right]_{j,N+1}
  = \left[\mathcal{X}\mathcal{M}^{(\alpha;\ell)}\right]_{j,\ell} \qquad \text{if } j\le N,
  \\
  &\left[ \widetilde{S}^{(\alpha)}_\mu\right]_{N+1,k}
  =\left[ \mathcal{M}^{(\alpha)}_\mu \right]_{N+1,k},
\end{align}
in terms of the expressions \eqref{XMbeta} and \eqref{XMell}.
Taking now the limit $\xi_{N+1}\to +\infty$ we obtain:
\begin{multline}
   \lim_{\xi_{N+1}\to +\infty} \mathcal{I}_{\{\xi_1,\ldots,\xi_{N+1}\},\{p_1,\ldots,p_N,\mu\}}[-\alpha\,  \tilde f_{P,Q}]
   =e^{-\mu-\sum_j (p_j-\xi_j)}
   \prod_{j=1}^N\frac{\sinh(\xi_j-\mu)}{\sinh(p_j-\mu)}
   \\
   \times
   (1+\alpha e^\eta )\, \frac{Q(\mu-\eta)\, P(\mu+i\pi)}{d(\mu)}\,\frac{\det_{N+1}\bar S_\mu^{(\alpha)}}{\det_N \mathcal{S}^{(0)}(\mathbf{q}\,|\, \mathbf{p})},
   \\
   =e^{-\mu-\sum_j (p_j-\xi_j)}\, (1+\alpha e^\eta )\,\frac{Q(\mu-\eta)\, P(\mu+i\pi)}{\prod_{j=1}^N\sinh(\mu-p_j)}\,\frac{\det_{N+1}\bar S_\mu^{(\alpha)}}{\det_N \mathcal{S}^{(0)}(\mathbf{q}\,|\, \mathbf{p})},
\end{multline}
in which
\begin{alignat}{2}
   &\left[ \bar S_\mu^{(\alpha)}\right]_{j,k}
   = \left[\mathcal{S}^{(\alpha)}(\mathbf{q}\,|\, \mathbf{p})\right]_{j,k}
   = \mathsf{S}_{Q,P}^{(\alpha)}(q_j,p_k),
   \qquad & &\text{if } j,k\le N, \label{Sbar-mu-1}\\
   &\left[ \bar S_\mu^{(\alpha)}\right]_{j,N+1}
   = \left[ \mathcal{S}^{(\alpha;\ell)}_{\mu}(\mathbf{q}\, |\, \mathbf{p})\right]_{j,\ell}
    \qquad & &\text{if } j\le N,\\
   &\left[ \bar S_\mu^{(\alpha)}\right]_{N+1,k}=\frac{e^{p_k} \, d(p_k)}{Q(p_k-\eta)\, P(p_k+i\pi)}
   \qquad & &\text{if } k\le N,\\
   &\left[ \bar S_\mu^{(\alpha)}\right]_{N+1,N+1}=\frac{e^{\mu} \, d(\mu)}{Q(\mu-\eta)\, P(\mu+i\pi)},
   \label{Sbar-mu-4}
\end{alignat}
in terms of the matrix of the scalar product \eqref{S_PQ-SP} and of the matrix elements \eqref{Sell-mu}.


Using this representation, we can rewrite the matrix element of the $D$-operator as
\begin{align}
 {}_\mathsf{n\!}\bra{P,\kappa,\epsilon}\,D(\mu)\,\ket{ Q,\kappa,\epsilon}_\mathsf{n}
  & =e^{-\sum_j (p_j-\xi_j)} \, 
  \frac{\det_{N+1}\widetilde{\mathcal S}^{(e^{-\eta})}(\mathbf{q}\,|\, \mathbf{p},\mu)}{\det_N  \mathcal{S}^{(0)}(\mathbf{q}\,|\, \mathbf{p})},       \label{mat-el-D-mu}
\end{align}
where the matrix $\widetilde{\mathcal S}^{(e^{-\eta})}(\mathbf{q}\,|\, \mathbf{p},\mu)$ is obtained from the matrix $\bar S_\mu^{(e^{-\eta})}$  \eqref{Sbar-mu-1}-\eqref{Sbar-mu-4} as
\begin{align}
   &\left[\widetilde{\mathcal S}^{(e^{-\eta})}(\mathbf{q}\, |\, \mathbf{p},\mu)\right]_{j,k}
   = \left[\bar S_\mu^{(e^{-\eta})}\right]_{j,k}
   \qquad  \text{for }\ k\le N,
   \label{Stilde1}\\
   &\left[\widetilde{\mathcal S}^{(e^{-\eta})}(\mathbf{q}\, |\, \mathbf{p},\mu)\right]_{j,N+1}
   =e^{-\mu}\, a(\mu)\,\frac{Q(\mu-\eta)\, P(\mu+i\pi)}{\prod_{j=1}^N\sinh(\mu-p_j)} 
   \left[ \bar S_\mu^{(e^{-\eta})}\right]_{j,N+1},
   \label{Stilde2}
\end{align}
for $1\le j\le N+1$.
Note that the whole dependence on $\mu$ of \eqref{mat-el-D-mu} is contained in the last column \eqref{Stilde2} of the determinant and that we have
\begin{equation}
    \left[\widetilde{\mathcal S}^{(e^{-\eta})}(\mathbf{q}\, |\, \mathbf{p},\xi_n-\eta)\right]_{j,N+1}
    =\frac{a(\xi_n)\,d(\xi_n-\eta)}{\tau_P(\xi_n)\,\tau_Q(\xi_n)}\,
      \left[\widetilde{\mathcal S}^{(e^{-\eta})}(\mathbf{q}\, |\, \mathbf{p},\xi_n)\right]_{j,N+1}.
\end{equation}
Hence, from \eqref{ff_s-1} and \eqref{ff_s+2}, we get
\begin{align}
   {}_\mathsf{n\!}\bra{P,\kappa,\epsilon}\,\sigma_n^-\,\ket{ Q,\kappa,\epsilon}_\mathsf{n}
   &={}_\mathsf{n\!}\bra{P,\kappa,\epsilon}\,\sigma_n^+\,\ket{ Q,\kappa,\epsilon}_\mathsf{n}
   \nonumber\\
   &=\epsilon\kappa \,     
    \frac{\prod_{k=1}^{n-1}\tau_P(\xi_k)}{\prod_{k=1}^n\tau_Q(\xi_k)}\, e^{-\sum_j (p_j-\xi_j)} \,
      \frac{ \det_{N+1}\widetilde{\mathcal S}^{(e^{-\eta})}(\mathbf{q}\,|\, \mathbf{p},\xi_n)}{\det_N  \mathcal{S}^{(0)}(\mathbf{q}\,|\, \mathbf{p})}.
      \label{ff-s-s+ext}
\end{align}

As previously, we can rewrite this result in terms of the determinant of the matrix of the scalar product plus a matrix of rank $1$. Expanding the determinant of $\widetilde{\mathcal S}^{(e^{-\eta})}(\mathbf{q}\,|\, \mathbf{p},\xi_n)$ with respect to the last row,  one can rewrite $\det_{N+1}\widetilde{\mathcal S}^{(e^{-\eta})}(\mathbf{q}\,|\, \mathbf{p},\xi_n)$ in \eqref{ff-s-s+ext} as
\begin{multline}\label{decomp-rank1}
   \det_{N+1}\widetilde{\mathcal S}^{(e^{-\eta})}(\mathbf{q}\,|\, \mathbf{p},\xi_n)
   =\left(\left[\widetilde{\mathcal S}^{(e^{-\eta})}(\mathbf{q}\, |\, \mathbf{p},\xi_n)\right]_{N+1,N+1}-1\right)
        \det_N\left[ \mathcal{S}^{(e^{-\eta})}(\mathbf{q}\,|\, \mathbf{p}) \right]
        \\
        + \det_N\left[ \mathcal{S}^{(e^{-\eta})}(\mathbf{q}\,|\, \mathbf{p}) 
           -\mathcal{P}^{(-)}(\mathbf{q}\,|\, \mathbf{p},\xi_n)\right] ,
\end{multline}
in which $ \mathcal{S}^{(e^{-\eta})}(\mathbf{q}\,|\, \mathbf{p})$ is the matrix of the scalar product \eqref{S_PQ-SP} with elements $\mathsf{S}_{Q,P}^{(e^{-\eta})}(q_j,p_k) $, $1\le j,k \le N$, and $\mathcal{P}^{(-)}(\mathbf{q}\,|\, \mathbf{p},\xi_n)$ is a matrix of rank 1 with elements
\begin{equation}
    \left[ \mathcal{P}^{(-)}(\mathbf{q}\,|\, \mathbf{p},\xi_n)\right]_{j,k}
    = \left[\widetilde{\mathcal S}^{(e^{-\eta})}(\mathbf{q}\, |\, \mathbf{p},\xi_n)\right]_{j,N+1}
               \left[\widetilde{\mathcal S}^{(e^{-\eta})}(\mathbf{q}\, |\, \mathbf{p},\xi_n)\right]_{N+1,k} .
\end{equation}
Noticing that
\begin{equation}
   \left[\widetilde{\mathcal S}^{(e^{-\eta})}(\mathbf{q}\, |\, \mathbf{p},\xi_n)\right]_{N+1,N+1}=0,
\end{equation}
one gets the result \eqref{ff-s-s+}-\eqref{mat-P-s-}.

Alternatively, one can use \eqref{fPQ-tau} before transforming \eqref{ratio-Imu} and multiply the determinant in the numerator of \eqref{ratio-Imu} by $\det_N\left[\mathcal{X}_\tau\right]$, where $\mathcal{X}_\tau$ is the matrix \eqref{X-tau}.
Then 
\begin{align}
   \det_{N+1}\mathcal{M}^{(\alpha)}_\mu
   =\frac{\det_{N+1}\widetilde{S}^{(\alpha)}_{\mu,\tau}}{\prod_{j=1}^N\tau_Q(\xi_j)\, \det_N[\mathcal{X}_\tau]}
\end{align}
with
\begin{align}
  &\left[ \widetilde{S}^{(\alpha)}_{\mu,\tau}\right]_{j,k}
  = \left[\mathcal{X}_\tau\mathcal{M}^{(\alpha)}_\tau\right]_{j,k}
  =\mathsf{S}^{(\alpha)}_{\tau_Q,\tau_P}(z_i,p_k) \qquad \text{if } j,k\le N,
  \\
  &\left[ \widetilde{S}^{(\alpha)}_{\mu,\tau}\right]_{j,N+1}
  =\mathsf{S}^{(\alpha)}_{\tau_Q,\tau_P}(z_i,\mu) \qquad \text{if } j\le N,
  \\
  &\left[ \widetilde{S}^{(\alpha)}_{\mu,\tau}\right]_{N+1,k}
  =\left[ \mathcal{M}^{(\alpha)}_\mu \right]_{N+1,k},
\end{align}
in terms of the expressions \eqref{matS-tau}. Note that the last row is unchanged.
Taking now the limit $\xi_{N+1}\to +\infty$ we obtain:
\begin{multline}
   \lim_{\xi_{N+1}\to +\infty} \mathcal{I}_{\{\xi_1,\ldots,\xi_{N+1}\},\{p_1,\ldots,p_N,\mu\}}[-\alpha\,  \tilde f_{P,Q}]
   =e^{-\mu-\sum_j (p_j-\xi_j)}
   \prod_{j=1}^N\frac{\sinh(\xi_j-\mu)}{\sinh(p_j-\mu)}
   \\
   \times
   \frac{1+\alpha e^\eta }{\prod_{j=1}^N\tau_Q(\xi_j)}\, \frac{\det_{N+1}\bar S_{\mu,\tau}^{(\alpha)}}{\det_N\mathcal{X}_\tau\,\det_N \mathcal{M}^{(0)}},
\end{multline}
in which
\begin{alignat}{2}
   &\left[ \bar S_{\mu,\tau}^{(\alpha)}\right]_{j,k}
   = \left[ \widetilde{S}^{(\alpha)}_{\mu,\tau}\right]_{j,k}
   =\mathsf{S}^{(\alpha)}_{\tau_Q,\tau_P}(z_i,p_k)
   \qquad & &\text{if } j,k\le N, \label{Sbar-mutau-1}\\
   &\left[ \bar S_{\mu,\tau}^{(\alpha)}\right]_{j,N+1}
   = \left[ \widetilde{S}^{(\alpha)}_{\mu,\tau}\right]_{j,N+1}
   =\mathsf{S}^{(\alpha)}_{\tau_Q,\tau_P}(z_i,\mu) 
    \qquad & &\text{if } j\le N,\\
   &\left[ \bar S_{\mu,\tau}^{(\alpha)}\right]_{N+1,k}=e^{p_k} 
   \qquad & &\text{if } k\le N,\\
   &\left[ \bar S_{\mu,\tau}^{(\alpha)}\right]_{N+1,N+1}=e^{\mu},
   \label{Sbar-mutau-4}
\end{alignat}
Using this representation, we can rewrite the matrix element of the $D$-operator as
\begin{align}
 {}_\mathsf{n\!}\bra{P,\kappa,\epsilon}\,D(\mu)\,\ket{ Q,\kappa,\epsilon}_\mathsf{n}
  & =e^{-\sum_j (p_j-\xi_j)} \, 
  \frac{\det_{N+1}\widetilde{\mathcal S}^{(e^{-\eta})}_{\tau_Q,\tau_P}(\mathbf{z}\,|\, \mathbf{p},\mu)}{\prod_{j=1}^N\tau_Q(\xi_j)\,\det_N\mathcal{X}_\tau\,\det_N \mathcal{M}^{(0)}},       \label{mat-el-D-mu-tau}
\end{align}
where the matrix $\widetilde{\mathcal S}^{(e^{-\eta})}_{\tau_Q,\tau_P}(\mathbf{z}\,|\, \mathbf{p},\mu)$ is obtained from the matrix $\bar S_{\mu,\tau}^{(e^{-\eta})}$  \eqref{Sbar-mutau-1}-\eqref{Sbar-mutau-4} as
\begin{align}
   &\left[\widetilde{\mathcal S}^{(e^{-\eta})}_{\tau_Q,\tau_P}(\mathbf{z}\,|\, \mathbf{p},\mu)\right]_{j,k}
   = \left[\bar S_{\mu,\tau}^{(e^{-\eta})}\right]_{j,k}
   \qquad  \text{for }\ k\le N,
   \label{Stildetau1}\\
   &\left[\widetilde{\mathcal S}^{(e^{-\eta})}_{\tau_Q,\tau_P}(\mathbf{z}\,|\, \mathbf{p},\mu)\right]_{j,N+1}
   =e^{-\mu}\, \frac{a(\mu)\, d(\mu)}{\prod_{j=1}^N\sinh(\mu-p_j)} 
   \left[ \bar S_{\mu,\tau}^{(e^{-\eta})}\right]_{j,N+1},
   \label{Stildetau2}
\end{align}
for $1\le j\le N+1$.
Note once again that the whole dependence on $\mu$ of \eqref{mat-el-D-mu-tau} is contained in the last column \eqref{Stildetau2} of the determinant and that we have
\begin{align}
   &\left[\widetilde{\mathcal S}^{(e^{-\eta})}_{\tau_Q,\tau_P}(\mathbf{z}\,|\, \mathbf{p},\xi_n)\right]_{j,N+1}
   =-\frac{a(\xi_n)\,\tau_Q(\xi_n)}{\prod_{\ell=1}^N\sinh(\xi_n-p_\ell)\, \sinh(z_j-\xi_n)},
   \\
   &\left[\widetilde{\mathcal S}^{(e^{-\eta})}_{\tau_Q,\tau_P}(\mathbf{z}\,|\, \mathbf{p},\xi_n-\eta)\right]_{j,N+1}
   =\frac{d(\xi_n-\eta)\, \tau_P(\xi_n)}{\prod_{\ell=1}^N\sin(\xi_n-p_\ell-\eta)\, \sinh(z_j-\xi_n)},
\end{align}
for $j\le N$, and
\begin{equation}
   \left[\widetilde{\mathcal S}^{(e^{-\eta})}_{\tau_Q,\tau_P}(\mathbf{z}\,|\, \mathbf{p},\xi_n)\right]_{N+1,N+1}
   =\left[\widetilde{\mathcal S}^{(e^{-\eta})}_{\tau_Q,\tau_P}(\mathbf{z}\,|\, \mathbf{p},\xi_n-\eta)\right]_{N+1,N+1}
   =0,
\end{equation}
so that, using \eqref{TQ-xi} for $P$, we can also explicitly check that
\begin{equation}
    \left[\widetilde{\mathcal S}^{(e^{-\eta})}_{\tau_Q,\tau_P}(\mathbf{z}\,|\, \mathbf{p},\xi_n-\eta)\right]_{j,N+1}
    =\frac{a(\xi_n)\,d(\xi_n-\eta)}{\tau_P(\xi_n)\,\tau_Q(\xi_n)}\,
      \left[\widetilde{\mathcal S}^{(e^{-\eta})}_{\tau_Q,\tau_P}(\mathbf{z}\,|\, \mathbf{p},\xi_n)\right]_{j,N+1}.
\end{equation}
This implies that
\begin{align}
   {}_\mathsf{n\!}\bra{P,\kappa,\epsilon}\,\sigma_n^-\,\ket{ Q,\kappa,\epsilon}_\mathsf{n}
   &={}_\mathsf{n\!}\bra{P,\kappa,\epsilon}\,\sigma_n^+\,\ket{ Q,\kappa,\epsilon}_\mathsf{n}
   \nonumber\\
   &\hspace{-2cm}=\epsilon\kappa \,     
    \frac{\prod_{k=1}^{n-1}\tau_P(\xi_k)}{\prod_{k=1}^n\tau_Q(\xi_k)}\, e^{-\sum_j (p_j-\xi_j)} \,
     \frac{\det_{N+1}\widetilde{\mathcal S}^{(e^{-\eta})}_{\tau_Q,\tau_P}(\mathbf{z}\,|\, \mathbf{p},\xi_n)}{\prod_{j=1}^N\tau_Q(\xi_j)\,\det_N\mathcal{X}_\tau\,\det_N \mathcal{M}^{(0)}}, 
      \label{ff-s-s+tau-ext}
\end{align}
Decomposing the determinant in the numerator of \eqref{ff-s-s+tau-ext} in the same way as in \eqref{decomp-rank1}, we obtain \eqref{ff-s-s+2}-\eqref{mat-P-s--2}.

\section{Conclusion}

In this article, we have explained how to compute the scalar products of a particular class of separate states, as well as the form factors of local operators, in the case of a model solvable by SoV and for which the Baxter $Q$-function is not a simple (trigonometric) polynomial. We have more precisely considered here the XXZ spin chain with antiperiodic boundary conditions: for this model, the $Q$-function solution of the (homogeneous) Baxter TQ-equation has double periodicity with respect to the functions defining the model, and the eigenstates are therefore not usual Bethe states.
Hence, the situation is quite different from the XXX case, so that the method proposed in \cite{KitMNT16} does not apply here.

We recall that, as usual, the scalar products and form factors can be expressed from SoV in terms of determinants with rows and columns labelled by the inhomogeneity parameters of the model, which is a priori not convenient for the consideration of the physical model at the homogeneous limit. We explained here how to transform such formulas into determinants in which rows and columns are labelled by the roots of the $Q$-functions, as in the periodic case solvable by Bethe Ansatz. Due to the difference of periodicity of the $Q$-function with respect to the periodic case, the obtained formulas are however different in their form from their periodic counterparts \cite{Sla89,KitMT99}. We also propose some new type of determinant representations, alternative to the aforementioned ones, and written in terms of the transfer matrix eigenvalues themselves. It is interesting to mention that the obtention of this new type of representation relies minimally on the form of the $Q$-function, and should therefore be quite easy to generalise in the context of other models solvable by SoV.

Of course, the representations written in Section~\ref{sec-results} are not the only ones that can be obtained by the method we propose. Different choices for the matrix $\mathcal{X}$ \eqref{mat-X} or $\mathcal{X}_\tau$ \eqref{X-tau} may lead to different final representations. Also, it is possible to play on the initial formulas obtained by SoV by using the fact that several expressions coincide when evaluated at the inhomogeneity parameters: this is indeed what we did in \eqref{fPQ-tau}, so as to transform \eqref{M-tau} instead of \eqref{M-beta}, and we may imagine several other variations at this level. However, the expressions we have obtained here seem to be among the most compact ones.
We therefore hope that these formulas, or some of their minor variations, can be useful for numerical studies, such as in \cite{BieKM02,BieKM03,CauM05,CauHM05}, or for analytical studies, such as in \cite{KitKMST09c,KitKMST11a}.

\bibliographystyle{SciPost_bibstyle}
\bibliography{/Users/vterras/Documents/Dropbox/Bib_files/biblio.bib}

\end{document}